\documentclass[11pt]{article}

\usepackage{amsmath}
\usepackage{amsfonts}
\usepackage{latexsym}
\usepackage{epsfig}
\usepackage{color}
\usepackage{feynmf}

\renewcommand{\d}{\mathrm{d}}

\newcommand{\dga}{{\dot{\alpha}}}

\newcommand{\tbgd}{{\tilde{\bar{\delta}}}}

\newcommand{\del}{{\partial}}
\newcommand{\bdel}{{\bar{\partial}}}

\newcommand{\DbD} [2]{\ensuremath{\frac{\bar{D}_{#1}^2 D_{#2}^2}{16}}}           % \bar{D} D over 16, two coordinates
\newcommand{\DDb} [2]{\ensuremath{\frac{D_{#1}^2 \bar{D}_{#2}^2}{16}}}           % D \bar{D} over 16, two coordinates
\newcommand{\dtw} [4]{\ensuremath{\widetilde{\delta}_{#2}^{\,#1}{}_{#3}{}^{#4}}} % delta twiddle, gauge indices up-down
\newcommand{\dtwup}[4]{\ensuremath{\widetilde{\delta}_{#2}^{\,#1}{}^{#3#4}}}     % delta twiddle, gauge indices up
\newcommand{\prop}[2]{\ensuremath{\frac{#1}{(\Box+\der_5^2)_#2}}}                % five-dim. propagator
\newcommand{\propsix}[2]{\ensuremath{\frac{#1}{(\Box+\del\bdel)_#2}}}            % five-dim. propagator
\newcommand{\sint}[1]{\ensuremath{\int \left(d^5x \, d^4\theta\right)_{#1}\;}}   % integration d5x d4theta
\newcommand{\Q}[2]{\ensuremath{Q^{#1}{}_{#2}}}                                   % Q, indices up-down
\newcommand{\fdu}[5]{\ensuremath{\frac{\delta J_{#1}{}_{#2}\,{}^{#4}}{\delta J_{#1}{}_{#3}\,{}^{#5}}}}
\newcommand{\Vol}{{\text{Vol}}}
\newcommand{\brkt}[2]{\left[ \begin{array}{c} {#1} \\ {#2} \end{array} \right]}

\DeclareMathSymbol{\mg}{\mathrel}{symbols}{"1D}

\newcommand{\ga}{\alpha}
\newcommand{\gb}{\beta}
\renewcommand{\gg}{\gamma}
\newcommand{\gd}{\delta}
\renewcommand{\ge}{\epsilon}

\newcommand{\gf}{\phi}
\newcommand{\gvf}{\varphi}

\newcommand{\gx}{\xi}
\newcommand{\gm}{\mu}

\newcommand{\gth}{\theta}
\newcommand{\gs}{\sigma}
\newcommand{\gt}{\tau}

\newcommand{\gp}{\pi}
\newcommand{\gps}{\psi}
\newcommand{\get}{\eta}

\newcommand{\gG}{\Gamma}
\newcommand{\gD}{\Delta}
\newcommand{\gF}{\Phi}

\newcommand{\gL}{\Lambda}
\newcommand{\gS}{\Sigma}
\newcommand{\gTh}{\Theta}

\newcommand{\cD}{{\cal D}}

\newcommand{\cI}{{\cal I}}
\newcommand{\cJ}{{\cal J}}

\newcommand{\cN}{{\cal N}}

\newcommand{\cS}{{\cal S}}

\newcommand{\tr}{\text{tr}}

\newcommand{\Id}{\text{\small 1}\hspace{-3.5pt}\text{1}}

\newcommand{\ra}{\rightarrow}

\renewcommand{\Im}{\text{Im}\ }

\newcommand{\der}{\partial}

\newcommand{\dsp}{\displaystyle}

\newcommand{\labl}[1]{\label{#1}}
\newcommand{\Kh}{K\"{a}hler}
\newcommand{\beq}{\begin{equation}}
\newcommand{\eeq}{\end{equation}}
\newcommand{\barr}{\begin{array}}
\newcommand{\earr}{\end{array}}
\newcommand{\equ}[1]{\begin{gather} #1 \end{gather}}
\newcommand{\equl}[1]{\begin{multline} #1 \end{multline}}
\newcommand{\equa}[1]{\begin{align} #1 \end{align}}

\newcommand{\tabu}[2]{\begin{tabular}{#1} #2 \end{tabular}}
\newcommand{\arry}[2]{\begin{array}{#1} #2 \end{array}}

\newcommand{\pmtrx}[1]{\begin{pmatrix} #1 \end{pmatrix}}

\newcommand{\non}{\nonumber}
\newcounter{oldcounter}

\newcommand{\bder}{\bar\partial}
\newcommand{\bn}{{\bar n}}

\newcommand{\bp}{{\bar p}}

\newcommand{\bw}{{\bar w}}

\newcommand{\bz}{{\bar z}}

\newcommand{\bC}{{\bar C}}
\newcommand{\bD}{{\bar D}}

\newcommand{\bF}{{\bar F}}

\newcommand{\bJ}{{\bar J}}

\newcommand{\bS}{{\bar S}}

\newcommand{\bU}{{\bar U}}

\newcommand{\bW}{{\bar W}}

\newcommand{\bZ}{{\bar Z}}
\newcommand{\bga}{{\bar \alpha}}
\newcommand{\bgb}{{\bar\beta}}

\newcommand{\bge}{{\bar\epsilon}}

\newcommand{\bgth}{{\bar\theta}}
\newcommand{\bgs}{{\bar\sigma}}

\newcommand{\bgF}{{\bar\Phi}}

\newcommand{\bgL}{{\bar\Lambda}}

\newcommand{\bgTh}{{\bar\Theta}}

\newcommand{\tgd}{{\tilde \delta}}

\newcommand{\Intr}{\mathbb{Z}}
\newcommand{\Cplx}{\mathbb{C}}
\newcommand{\Real}{\mathbb{R}}

\newcommand{\ba}[2]{\[\begin{array}{#2}\label{#1}}
\newcommand{\ea}{\end{array}\]}
\newcommand{\be}{\begin{equation}}
\newcommand{\ee}{\end{equation}}
\newcommand{\bea}{\begin{eqnarray}}
\newcommand{\eea}{\end{eqnarray}}

\newcommand{\U}[1]{\mathrm{U(#1)}}
\newcommand{\SU}[1]{\mathrm{SU(#1)}}

\newcommand{\rep}[1]{\mathbf{#1}}
\newcommand{\crep}[1]{\overline{\rep{#1}}}

\newcommand{\captn}[1]{\vspace{-3ex}\caption{\small #1}}

\newcommand{\removed}[1]{}

\newcommand{\Propagators}{
\begin{figure}
\begin{center}
\tabu{cccc}{
      \makebox[80pt]{   \begin{fmfgraph*}(50,20)     
      \fmfleft{i}
      \fmflabel{$\mathbf{\bgF_\pm}$}{i}
      \fmfright{o}
      \fmflabel{$\mathbf{\gF_\pm}$}{o}
      \fmf{plain}{i,o}
    \end{fmfgraph*}
} 
\qquad & \qquad 
      \makebox[80pt]{    \begin{fmfgraph*}(50,20)     %%% V-[V]-V loop %%%
      \fmfleft{i}
      \fmflabel{$\mathbf{\gF_+}$}{i}
      \fmfright{o}
      \fmflabel{$\mathbf{\gF_-}$}{o}
      \fmf{plain_arrow}{i,o}
    \end{fmfgraph*}
} 
\qquad & \qquad 
      \makebox[80pt]{    \begin{fmfgraph*}(50,20)     %%% V-[V/S]-V loop %%%
      \fmfleft{i}
      \fmflabel{$\mathbf{V}$}{i}
      \fmfright{o}
      \fmflabel{$\mathbf{V}$}{o}
      \fmf{boson}{i,o}
    \end{fmfgraph*}
} 
\qquad & \qquad  
      \makebox[80pt]{   \begin{fmfgraph*}(50,20)     %%% V-[C]-V loop %%%
      \fmfleft{i}
      \fmflabel{$\mathbf{\bS}$}{i}
      \fmfright{o}
      \fmflabel{$\mathbf{S}$}{o}
      \fmf{dashes,tension=0.4}{i,o}
    \end{fmfgraph*}
}
}
\\
\tabu{cc}{
      \makebox[80pt]{   \begin{fmfgraph*}(50,20)     
      \fmfleft{i}
      \fmflabel{$\mathbf{\bC'}$}{i}
      \fmfright{o}
      \fmflabel{$\mathbf{C}$}{o}
      \fmf{dots,tension=0.4}{i,o}
    \end{fmfgraph*}
} 
\qquad  &  \qquad  
      \makebox[80pt]{    \begin{fmfgraph*}(50,20)     %%% V-[V]-V loop %%%
      \fmfleft{i}
      \fmflabel{$\mathbf{C'}$}{i}
      \fmfright{o}
      \fmflabel{$\mathbf{\bC}$}{o}
      \fmf{dots,tension=0.4}{i,o}
    \end{fmfgraph*}
} 
} 
\end{center}
\captn{\label{fg:Propagators}
This picture gives our drawing conventions for the propagators
which we employ throughout this paper.  In particular, there are two
chiral multiplet propagators: The first one corresponds to the
diagonal terms in  \eqref{propH}, while the second refers to the
off--diagonal parts. 
}
\end{figure}
}

\newcommand{\HyperVert}{
\begin{figure}[t]
\begin{center}
\tabu{ccc}{
\makebox[100pt]{    \begin{fmfgraph*}(50,50) %%% S_H:  V  and 2 hypers (+)
      \fmfleft{i}
      \fmflabel{$\mathbf{V}$}{i}
      \fmfright{o1,o2,o3,o4}
      \fmflabel{$\mathbf{\bgF_\pm}$}{o2}
      \fmflabel{$\mathbf{\gF_\pm}$}{o3}
      \fmf{boson, tension=2}{i,v1}
      \fmf{plain}{o2,v1}
      \fmf{plain}{v1,o3}
      \fmfdotn{v}{1}
    \end{fmfgraph*}
} 
\quad & \quad  
\makebox[100pt]{    \begin{fmfgraph*}(50,50) %%% S_H: 2Vs and 2 hypers (+)
      \fmfleft{i1,i2,i3,i4}
      \fmflabel{$\mathbf{V\,\,}$}{i2}
      \fmflabel{$\mathbf{\bgF_\pm}$}{i3}
      \fmfright{o1,o2,o3,o4}
      \fmflabel{$\mathbf{\,\,V}$}{o2}
      \fmflabel{$\mathbf{\gF_\pm}$}{o3}
      \fmf{boson}{i2,v1}
      \fmf{plain}{i3,v1}
      \fmf{boson}{v1,o2}
      \fmf{plain}{v1,o3}
      \fmfdotn{v}{1}
    \end{fmfgraph*}
} 
\quad & \quad  
\makebox[100pt]{    \begin{fmfgraph*}(50,50) %%% S_H:  S  and 2 hypers (+-)
      \fmfleft{i}
      \fmflabel{$\mathbf{S}$}{i}
      \fmfright{o1,o2,o3,o4}
      \fmflabel{$\mathbf{\gF_-}$}{o2}
    \fmflabel{$\mathbf{\gF_+}$}{o3}
      \fmf{dashes, tension=2}{i,v1}
      \fmf{plain}{v1,o2}
      \fmf{plain}{v1,o3}
      \fmfdotn{v}{1}
    \end{fmfgraph*}
}
}
\end{center}
\captn{\labl{fg:SH_vertices}
These interaction vertices involve the coupling of the gauge
superfields $V$ and $S$ to the hyper multiplet chiral superfields
$\gF_+$ and $\gF_-$.}
\end{figure}
}

\newcommand{\VectorVert}{
\begin{figure}[t]
\begin{center}
\tabu{ccc}{ 
\makebox[100pt]{    \begin{fmfgraph*}(50,50) %%% S_H:  3Vs
      \fmfleft{i}
      \fmflabel{$\mathbf{V}$}{i}
      \fmfright{o1,o2,o3,o4}
      \fmflabel{$\mathbf{V}$}{o2}
      \fmflabel{$\mathbf{V}$}{o3}
      \fmf{boson, tension=2}{i,v1}
      \fmf{boson}{v1,o2}
      \fmf{boson}{v1,o3}
      \fmfdotn{v}{1}
    \end{fmfgraph*}
} 
\quad & \quad 
\makebox[100pt]{    \begin{fmfgraph*}(50,50) %%% S_H:  S  and 2 Vs
      \fmfleft{i}
      \fmflabel{$\mathbf{S}$}{i}
      \fmfright{o1,o2,o3,o4}
      \fmflabel{$\mathbf{V}$}{o2}
      \fmflabel{$\mathbf{V}$}{o3}
      \fmf{dashes, tension=2}{i,v1}
      \fmf{boson}{v1,o2}
      \fmf{boson}{v1,o3}
      \fmfdotn{v}{1}
    \end{fmfgraph*}
} 
\quad & \quad 
\makebox[100pt]{    \begin{fmfgraph*}(50,50) %%% S_H:  V  and S and bS
      \fmfleft{i}
      \fmflabel{$\mathbf{V}$}{i}
      \fmfright{o1,o2,o3,o4}
      \fmflabel{$\mathbf{S}$}{o2}
      \fmflabel{$\mathbf{\bS}$}{o3}
      \fmf{boson, tension=2}{i,v1}
      \fmf{dashes}{v1,o2}
      \fmf{dashes}{v1,o3}
      \fmfdotn{v}{1}
    \end{fmfgraph*}
} 
\\[0ex]
\makebox[100pt]{    \begin{fmfgraph*}(50,50) %%% 4Vs
      \fmfleft{i1,i2,i3,i4}
      \fmflabel{$\mathbf{V}$}{i2}
      \fmflabel{$\mathbf{V}$}{i3}
      \fmfright{o1,o2,o3,o4}
      \fmflabel{$\mathbf{V}$}{o2}
      \fmflabel{$\mathbf{V}$}{o3}
      \fmf{boson}{i2,v1}
      \fmf{boson}{i3,v1}
      \fmf{boson}{v1,o2}
      \fmf{boson}{v1,o3}
      \fmfdotn{v}{1}
    \end{fmfgraph*}
} 
\quad & \quad 
\makebox[100pt]{    \begin{fmfgraph*}(50,50) %%% 2Vs and S and bS
      \fmfleft{i1,i2,i3,i4}
      \fmflabel{$\mathbf{V}$}{i2}
      \fmflabel{$\mathbf{S}$}{i3}
      \fmfright{o1,o2,o3,o4}
      \fmflabel{$\mathbf{V}$}{o2}
      \fmflabel{$\mathbf{\bS}$}{o3}
      \fmf{boson}{i2,v1}
      \fmf{dashes}{i3,v1}
      \fmf{boson}{v1,o2}
      \fmf{dashes}{v1,o3}
      \fmfdotn{v}{1}
    \end{fmfgraph*}
} 
\quad & \quad 
\makebox[100pt]{    \begin{fmfgraph*}(50,50) %%% 3Vs und S
      \fmfleft{i1,i2,i3,i4}
      \fmflabel{$\mathbf{V}$}{i2}
      \fmflabel{$\mathbf{V}$}{i3}
      \fmfright{o1,o2,o3,o4}
      \fmflabel{$\mathbf{V}$}{o2}
      \fmflabel{$\mathbf{S}$}{o3}
      \fmf{boson}{i2,v1}
      \fmf{boson}{i3,v1}
      \fmf{boson}{v1,o2}
      \fmf{dashes}{v1,o3}
      \fmfdotn{v}{1}
    \end{fmfgraph*}
} 
}
\end{center}
\captn{\label{fg:SV_vertices}
These vertices encode the self interactions of the gauge 
multiplet involving the vector superfield $V$ and the chiral
superfield $S$. 
}
\end{figure}
}

\newcommand{\GhostVert}{
\begin{figure}[t]
\begin{center}
\makebox[100pt]{    \begin{fmfgraph*}(50,50) %%% S_Gh:  V  and 2Cs
      \fmfleft{i}
      \fmflabel{$\mathbf{V}$}{i}
      \fmfright{o1,o2,o3,o4}
      \fmflabel{$\mathbf{C'}$}{o2}
      \fmflabel{$\mathbf{C}$}{o3}
      \fmf{boson, tension=2}{i,v1}
      \fmf{dots}{v1,o2}
      \fmf{dots}{v1,o3}
      \fmfdotn{v}{1}
    \end{fmfgraph*}
} 
\qquad 
\makebox[100pt]{    \begin{fmfgraph*}(50,50) %%% S_Gh:  V  and 2Cs
      \fmfleft{i}
      \fmflabel{$\mathbf{V}$}{i}
      \fmfright{o1,o2,o3,o4}
      \fmflabel{$\mathbf{\bC'}$}{o2}
      \fmflabel{$\mathbf{C}$}{o3}
      \fmf{boson, tension=2}{i,v1}
      \fmf{dots}{v1,o2}
      \fmf{dots}{v1,o3}
      \fmfdotn{v}{1}
    \end{fmfgraph*}
} 
\qquad 
\makebox[100pt]{    \begin{fmfgraph*}(50,50) %%% S_Gh:  S and 2Cs
      \fmfleft{i}
      \fmflabel{$\mathbf{S}$}{i}
      \fmfright{o1,o2,o3,o4}
      \fmflabel{$\mathbf{C}$}{o2}
      \fmflabel{$\mathbf{\bC'}$}{o3}
      \fmf{dashes, tension=2}{i,v1}
      \fmf{dots}{v1,o2}
      \fmf{dots}{v1,o3}
      \fmfdotn{v}{1}
    \end{fmfgraph*}
} 
\\[2ex]
\makebox[100pt]{    \begin{fmfgraph*}(50,50) %%% S_Gh:  2Vs and 2Cs
      \fmfleft{i1,i2,i3,i4}
      \fmflabel{$\mathbf{V}$}{i2}
      \fmflabel{$\mathbf{C}$}{i3}
      \fmfright{o1,o2,o3,o4}
      \fmflabel{$\mathbf{V}$}{o2}
      \fmflabel{$\mathbf{C'}$}{o3}
      \fmf{boson}{i2,v1}
      \fmf{dots}{i3,v1}
      \fmf{boson}{v1,o2}
      \fmf{dots}{v1,o3}
      \fmfdotn{v}{1}
    \end{fmfgraph*}
} 
\qquad 
\makebox[100pt]{    \begin{fmfgraph*}(50,50) %%% S_Gh:  2Vs and 2Cs
      \fmfleft{i1,i2,i3,i4}
      \fmflabel{$\mathbf{V}$}{i2}
      \fmflabel{$\mathbf{C}$}{i3}
      \fmfright{o1,o2,o3,o4}
      \fmflabel{$\mathbf{V}$}{o2}
      \fmflabel{$\mathbf{\bC'}$}{o3}
      \fmf{boson}{i2,v1}
      \fmf{dots}{i3,v1}
      \fmf{boson}{v1,o2}
      \fmf{dots}{v1,o3}
      \fmfdotn{v}{1}
    \end{fmfgraph*}
} 
\end{center}
  \captn{\labl{fg:SGh_vertices}
The ghosts $C$ and $C'$ only interact with the vector multiplet
superfields  $V$ and $S$. 
}
\end{figure}
}

\newcommand{\VVSelfEnergyHypers}{
\begin{figure}
\begin{center}
\tabu{ccc}{
    \makebox[100pt]{     \begin{fmfgraph*}(75,50) %%% T_{++} loop %%%
      \fmfleft{i}
      \fmflabel{$\mathbf{V}$}{i}
      \fmfright{o}
      \fmflabel{$\mathbf{V}$}{o}
      \fmf{boson}{i,v1}
      \fmf{plain,left,tension=0.4,label=$\mathbf{\gF_\pm}$}{v1,v2}
      \fmf{plain,left,tension=0.4,label=$\mathbf{\gF_\pm}$}{v2,v1}
      \fmf{boson}{v2,o}
      \fmfdotn{v}{2}
    \end{fmfgraph*}
} 
\qquad & \qquad 
    \makebox[100pt]{    \begin{fmfgraph*}(75,50)     %%% T_{--} loop %%%
      \fmfleft{i}
      \fmflabel{$\mathbf{V}$}{i}
      \fmfright{o}
      \fmflabel{$\mathbf{V}$}{o}
      \fmf{boson}{i,v1}
      \fmf{plain_arrow,left,tension=0.4,label=$\mathbf{\gF_-}$}{v1,v2}
      \fmf{plain_arrow,left,tension=0.4,label=$\mathbf{\gF_+}$}{v2,v1}
      \fmf{boson}{v2,o}
      \fmfdotn{v}{2}
    \end{fmfgraph*}
} 
\qquad & \qquad 
      \makebox[100pt]{    \begin{fmfgraph*}(75,50) %%% T_+ tadpole %%%
      \fmfbottom{i,o}
      \fmflabel{$\mathbf{V}$}{i}
       \fmflabel{$\mathbf{V}$}{o}
      \fmf{boson}{i,v1}
      \fmf{plain,tension=0.7,label=$\mathbf{\gF_\pm}$}{v1,v1}
      \fmf{boson}{v1,o}
      \fmfdotn{v}{1}
    \end{fmfgraph*}
}
\\[3ex]
A$_\pm$ \qquad &\qquad  B \qquad &\qquad  C$_\pm$   
\\[-3ex] 
}
\end{center}
\caption{\label{fg:VV_from_hypers}
The gauge multiplet receives $VV$ self energy corrections from the
hyper multiplet.  The proper self energy graphs are labeled
\ref{fg:VV_from_hypers}.A$_\pm$ and \ref{fg:VV_from_hypers}.B. 
The tadpole graph is denoted by \ref{fg:VV_from_hypers}.C$_\pm$. 
}
\end{figure}
}

\newcommand{\VVSelfEnergyVS}{
\begin{figure}
\begin{center}
\tabu{cccc}{
      \makebox[100pt]{   \begin{fmfgraph*}(75,50)     %%% V-[S]-V loop %%%
      \fmfleft{i}
      \fmflabel{$\mathbf{V}$}{i}
      \fmfright{o}
      \fmflabel{$\mathbf{V}$}{o}
      \fmf{boson}{i,v1}
      \fmf{dashes,left,tension=0.4,label=$\mathbf{S}$}{v1,v2}
      \fmf{dashes,left,tension=0.4,label=$\mathbf{S}$}{v2,v1}
      \fmf{boson}{v2,o}
      \fmfdotn{v}{2}
    \end{fmfgraph*}
} 
~&~
      \makebox[100pt]{    \begin{fmfgraph*}(75,50)     %%% V-[V]-V loop %%%
      \fmfleft{i}
      \fmflabel{$\mathbf{V}$}{i}
      \fmfright{o}
      \fmflabel{$\mathbf{V}$}{o}
      \fmf{boson}{i,v1}
      \fmf{boson,left,tension=0.4,label=$\mathbf{V}$}{v1,v2}
      \fmf{boson,left,tension=0.4,label=$\mathbf{V}$}{v2,v1}
      \fmf{boson}{v2,o}
      \fmfdotn{v}{2}
    \end{fmfgraph*}
} 
~&~
      \makebox[100pt]{    \begin{fmfgraph*}(75,50)     %%% V-[V/S]-V loop %%%
      \fmfleft{i}
      \fmflabel{$\mathbf{V}$}{i}
      \fmfright{o}
      \fmflabel{$\mathbf{V}$}{o}
      \fmf{boson}{i,v1}
      \fmf{boson,left,tension=0.4,label=$\mathbf{V}$}{v1,v2}
      \fmf{dashes,left,tension=0.4,label=$\mathbf{S}$}{v2,v1}
      \fmf{boson}{v2,o}
      \fmfdotn{v}{2}
    \end{fmfgraph*}
} 
~&~ 
      \makebox[100pt]{   \begin{fmfgraph*}(75,50)     %%% V-[C]-V loop %%%
      \fmfleft{i}
      \fmflabel{$\mathbf{V}$}{i}
      \fmfright{o}
      \fmflabel{$\mathbf{V}$}{o}
      \fmf{boson}{i,v1}
      \fmf{dots,left,tension=0.4,label=$\mathbf{C}$}{v1,v2}
      \fmf{dots,left,tension=0.4,label=$\mathbf{C}$}{v2,v1}
      \fmf{boson}{v2,o}
      \fmfdotn{v}{2}
    \end{fmfgraph*}
}
\\[2ex] 
A ~&~ B ~&~ C ~&~ D
} 
\\[6ex]  
\tabu{ccc}{
    \makebox[100pt]{    \begin{fmfgraph*}(75,50)     %%% V-[V]-V tadpole %%%
      \fmfleft{i}
      \fmflabel{$\mathbf{V}$}{i}
      \fmfright{o}
      \fmflabel{$\mathbf{V}$}{o}
      \fmf{boson}{i,v1}
      \fmf{boson,right,tension=0.9,label=$\mathbf{V}$}{v1,v1}
      \fmf{boson}{v1,o}
      \fmfdotn{v}{1}
    \end{fmfgraph*}
} 
\qquad & \qquad  
    \makebox[100pt]{    \begin{fmfgraph*}(75,50)     %%% V-[S]-V tadpole %%%
      \fmfleft{i}
      \fmflabel{$\mathbf{V}$}{i}
      \fmfright{o}
      \fmflabel{$\mathbf{V}$}{o}
      \fmf{boson}{i,v1}
      \fmf{dashes,right,tension=0.9,label=$\mathbf{S}$}{v1,v1}
      \fmf{boson}{v1,o}
      \fmfdotn{v}{1}
    \end{fmfgraph*}
} 
\qquad & \qquad 
    \makebox[100pt]{   \begin{fmfgraph*}(75,50)     %%% V-[C]-V tadpole %%%
      \fmfleft{i}
      \fmflabel{$\mathbf{V}$}{i}
      \fmfright{o}
      \fmflabel{$\mathbf{V}$}{o}
      \fmf{boson}{i,v1}
      \fmf{dots,right,tension=0.9,label=$\mathbf{C}$}{v1,v1}
      \fmf{boson}{v1,o}
      \fmfdotn{v}{1}
    \end{fmfgraph*}
}
\\[-3ex]
E \qquad &\qquad F \qquad &\qquad G 
}
\end{center}
\captn{\label{fg:VV_from_VSC}
The gauge contributions to the $VV$ part of the gauge multiplet self
energy are due to the $V$self coupling, the interactions with the chiral
superfield  $S$ and the ghost superfields $C$ and $C'$. In the first
line the genuine self energy graphs are labeled 
\ref{fg:VV_from_VSC}.A to \ref{fg:VV_from_VSC}.D. 
The tadpole graphs on the second line are referred to as 
\ref{fg:VV_from_VSC}.E to \ref{fg:VV_from_VSC}.G. 
}
\end{figure}
}

\newcommand{\SSSelfEnergyVS}{
\begin{figure}
\begin{center}
\tabu{cccc}{
    \makebox[100pt]{     \begin{fmfgraph*}(75,50) 
      \fmfleft{i}
      \fmflabel{$\mathbf{\bS}$}{i}
      \fmfright{o}
      \fmflabel{$\mathbf{S}$}{o}
      \fmf{dashes,tension=0.4}{i,v1}
      \fmf{boson,left,tension=0.16,label=$\mathbf{V}$}{v1,v2}
      \fmf{dashes,left,tension=0.16,label=$\mathbf{S}$}{v2,v1}
      \fmf{dashes,tension=0.4}{v2,o}
      \fmfdotn{v}{2}
    \end{fmfgraph*}
} 
\qquad & \qquad 
    \makebox[100pt]{    \begin{fmfgraph*}(75,50)    
      \fmfleft{i}
      \fmflabel{$\mathbf{V}$}{i}
      \fmfright{o}
      \fmflabel{$\mathbf{S}$}{o}
      \fmf{boson,tension=0.4}{i,v1}
      \fmf{boson,left,tension=0.16,label=$\mathbf{V}$}{v1,v2}
      \fmf{dashes,right,tension=0.16,label=$\mathbf{S}$}{v1,v2}
      \fmf{dashes,tension=0.4}{v2,o}
      \fmfdotn{v}{2}
    \end{fmfgraph*}
} 
\qquad & \qquad 
    \makebox[100pt]{    \begin{fmfgraph*}(75,50)    
      \fmfleft{i}
      \fmflabel{$\mathbf{V}$}{i}
      \fmfright{o}
      \fmflabel{$\mathbf{S}$}{o}
      \fmf{boson,tension=0.4}{i,v1}
      \fmf{dots,left,tension=0.16,label=$\mathbf{C}$}{v1,v2}
      \fmf{dots,right,tension=0.16,label=$\mathbf{C}$}{v1,v2}
      \fmf{dashes,tension=0.4}{v2,o}
      \fmfdotn{v}{2}
    \end{fmfgraph*}
} 
\\[3ex]
A \qquad & \qquad B \qquad & \qquad C 
\\[-3ex] 
}
\end{center}
\caption{\label{fg:SS_from_VS}
The $S \bS$ self energy graph is given in figure \ref{fg:SS_from_VS}.A.
The mixing between the 4D superfields $V$ and $S$ corresponding to 
the third term of \eqref{NA5DSV2} is renormalized by the diagrams
\ref{fg:SS_from_VS}.B and \ref{fg:SS_from_VS}.C. 
The last diagram has ghosts in the loop. 
}
\end{figure}
}

\newcommand{\SSSelfEnergyHypers}{
\begin{figure}
\begin{center}
\tabu{ccc}{
    \makebox[100pt]{     \begin{fmfgraph*}(75,50) 
      \fmfleft{i}
      \fmflabel{$\mathbf{\bS}$}{i}
      \fmfright{o}
      \fmflabel{$\mathbf{S}$}{o}
      \fmf{dashes,tension=0.4}{i,v1}
      \fmf{plain,left,tension=0.14,label=$\mathbf{\gF_\pm}$}{v1,v2}
      \fmf{plain,left,tension=0.14,label=$\mathbf{\gF_\pm}$}{v2,v1}
      \fmf{dashes,tension=0.4}{v2,o}
      \fmfdotn{v}{2}
    \end{fmfgraph*}
} 
\qquad \qquad & \qquad \qquad 
    \makebox[100pt]{     \begin{fmfgraph*}(75,50) 
      \fmfleft{i}
      \fmflabel{$\mathbf{V}$}{i}
      \fmfright{o}
      \fmflabel{$\mathbf{S}$}{o}
      \fmf{boson,tension=0.4}{i,v1}
      \fmf{plain,left,tension=0.14,label=$\mathbf{\gF_\pm}$}{v1,v2}
      \fmf{plain_arrow,right,tension=0.14,label=$\mathbf{\gF_\pm}$}{v1,v2}
      \fmf{dashes,tension=0.4}{v2,o}
      \fmfdotn{v}{2}
    \end{fmfgraph*}
} 
\\[3ex]
A$_\pm$ \qquad \qquad  & \qquad \qquad  B$_\pm$
\\[-3ex] 
}
\end{center}
\caption{\label{fg:SS_from_Hypers}
The gauge multiplet receives $\bS S$ self energy corrections from the
hyper multiplet as is depicted in figure
\ref{fg:SS_from_Hypers}.A$_\pm$. 
In addition the hyper multiplet gives rise to mixing between the
4D superfields $V$ and $S$, see  
\ref{fg:SS_from_Hypers}.B$_\pm$.   
}
\end{figure}
}

\begin{document}
  \begin{fmffile}{gfx}
    \fmfpen{thin}

 \thispagestyle{empty}

\begin{flushright}
SIAS-CMTP-06-1\\ 
USTC-ICTS-06-01 \\ 
hep-th/0602155
\end{flushright}
\vskip 2 cm
\begin{center}
{\Large {\bf 
Quantum Corrections to Non-Abelian SUSY Theories on Orbifolds
}
}
\\[0pt]
\vspace{1.23cm}
{\large
{\bf Stefan Groot Nibbelink$^{a,}$\footnote{
{{ {\ {\ {\ E-mail: nibbelin@ustc.edu.cn}}}}}}}, 
{\bf Mark Hillenbach$^{b,}$\footnote{
{{ {\ {\ {\ E-mail: mark@th.physik.uni-bonn.de}}}}}}}
\bigskip }\\[0pt]
\vspace{0.23cm}
${}^a$ {\it 
Shanghai Institute for Advanced Study, 
University of Science and Technology of China,  \\ 
99 Xiupu Rd, Pudong, Shanghai 201315, P.R.\ China
\\[1ex] 
Interdisciplinary Center for Theoretical Study, 
University of Science and Technology of China,  \\
Hefei, Anhui 230026, P.R.\ China 
 \\}
\vspace{0.23cm}
${}^b$ {\it 
Physikalisches Institut, Universit\"at Bonn, 
Nussallee 12, D-53115 Bonn, Germany
\\}
\bigskip
\vspace{1.4cm} 
\end{center}
\subsection*{\centering Abstract}

We consider supersymmetric non-Abelian gauge theories coupled to hyper
multiplets on five and six dimensional orbifolds, $S^1/\Intr_2$ and
$T^2/\Intr_N$, respectively. We compute the bulk and local fixed  
point renormalizations of the gauge couplings. To this end we extend
supergraph techniques to these orbifolds by defining orbifold
compatible delta functions. We develop their properties in detail. To 
cancel the bulk one-loop divergences the bulk gauge kinetic terms and
dimension six higher derivative operators are required. The gauge
couplings renormalize at the $\Intr_N$ fixed points due to vector
multiplet self interactions; the hyper multiplet renormalizes only
non-$\Intr_2$ fixed points. In 6D the Wess-Zumino-Witten term and a
higher derivative analogue have to renormalize in the bulk as well to
preserve 6D gauge invariance.

\newpage 

%%% Local Variables: 
%%% mode: latex
%%% TeX-master: "paper"
%%% End: 

 \setcounter{page}{1}

%  \tableofcontents

\section{Introduction}
\labl{sc:intro}

Theories of extra dimensions have been investigated for a long time 
after the pioneering work by Kaluza and Klein. More recently, string
theory has been very important to stimulate research into this
direction, because the natural number of dimensions for string theory
seems to be ten. Not only in the string theory community the topic of
extra dimensions has attracted a lot of attention, also
phenomenologists looked at this possibility. This was initiated by the
papers  \cite{Arkani-Hamed:1998rs,Antoniadis:1998ig}. Most of the
phenomenological activity has focused on five dimensional (5D) models,
in particular models on simple 1D orbifolds like $S^1/\Intr_2$ or
$S^1/\Intr_2\times \Intr_2'$
\cite{Mirabelli:1997aj,Barbieri:2000vh,Delgado:1998qr}. One of the
main reasons to turn to orbifolds is that they naturally lead to
chiral fermions. And if the extra dimensional theory is supersymmetric
then only by orbifolding 4D $\cN=1$ supersymmetry  can  be
recovered. Also 2D orbifolds like $T^2/\Intr_N$ have been considered
in phenomenological applications in 6D. To obtain phenomenological
models from (heterotic) string theory one often uses 6D orbifolds. In
this paper we will focus primarily on orbifolds in 5D and 6D, but
these results can be easily extended to the 10D string theory and 11D
M-theory settings.

There have been many investigations of quantum corrections to field
theories on orbifolds. An issue that received particular attention is
the generation of the Fayet-Iliopoulos terms at the fixed points of 1D
orbifolds
\cite{Ghilencea:2001bw,Barbieri:2001cz,GrootNibbelink:2002wv,Barbieri:2002ic,GrootNibbelink:2002qp,Marti:2002ar}.
Another issue of investigations in extra dimensions is the question of the
running of the gauge coupling and possible gauge
coupling unification. Initial investigations like \cite{Dienes:1998vh,Dienes:1998vg} took a naive approach to this problem,
but it was soon widely accepted that this running is in principle sensitive
to the ultra--violet (UV) completion of the theory
\cite{Hebecker:2002vm}.

There are various issues that one has to be aware of when discussing
5D and 6D theories. In 6D the constraints of anomalies
are very severe \cite{Seiberg:1996qx,Berkooz:1996iz,Danielsson:1997kt}. 
This means that one has to be very careful when one tries to obtain a
consistent theory in a bottom-up approach. However, if one considers
(heterotic) string compactifications it is essentially guaranteed that no dangerous
anomalies can ever arise. In the present work the issue of anomalies
is not so important, because we are simply interested in the
corrections to bulk and brane gauge operators due to various
supermultiplets. These results can then be in particular be applied to 
anomaly free models in 6D.

In a previous publication \cite{GrootNibbelink:2005vi} we studied the
gauge coupling running by calculating the photon self-energy in extra
dimensions. We focused on the renormalization of the gauge operators in
Abelian supersymmetric field theories on 5D and 6D orbifolds. In
general one would expect that both bulk and fixed point gauge
couplings would renormalize
\cite{Georgi:2000ks,vonGersdorff:2003dt,GrootNibbelink:2003gd}, 
but we found that for a charged bulk hyper multiplet the contributions
cancel out at the fixed points of $S^1/\Intr_2$. However, for a 6D orbifold
$T^2/\Intr_N$ this cancellation does not persist, except for the fixed
points that are invariant under a $\Intr_2$ symmetry. The other feature we found
is that in the 6D case also a dimension six higher derivative term for the
gauge multiplet is required to cancel all divergences. The observation
that such higher derivative operators are generated is no surprise, it is
simply one of the consequences that we consider 6D theories which are non-renormalizable.
%%% I removed ``NOT'' in this sentence and thus changed
%%% its meaning, please observe if correct!

Such higher derivative operators have also been found recently for other
quantities in theories of extra dimensions: In a 6D supersymmetric
model compactified on $T^2/\Intr_2$ such operators were obtained in
the context of Scherk-Schwarz supersymmetry breaking 
\cite{Ghilencea:2005nt}. And even in 5D orbifold models they can arise
if brane localized interactions for bulk fields are considered
\cite{Ghilencea:2004sq}. This can be understood in the
Kaluza-Klein mode picture by realizing that for brane localized
interactions, the Kaluza-Klein number is not conserved, so that double
sums can arise at one loop, raising the degree of divergence of the
corresponding sum/integral. Such higher derivative theories may have
remarkable UV properties, and might actually be renormalizable, as
speculated in  \cite{Smilga:2004cy,Ivanov:2005qf,Ivanov:2005kz}.

In this work we extend and complete the work that was initiated in 
our previous paper \cite{GrootNibbelink:2005vi} to include non-Abelian
gauge interactions. We employ again the method of
representing 5D and 6D SUSY theories by  $\cN=1$ 4D superfields
\cite{Arkani-Hamed:2001tb,Hebecker:2001ke,Marti:2001iw,Dudas:2004ni} and give a
detailed account of how to apply supergraph techniques to 5D and 6D
orbifolds. (For applications of supergraph techniques in the context of
supergravity see \cite{Buchbinder:2003qu}.) While the renormalization of the gauge couplings due to the hyper
multiplets is a straightforward extension of the previous work, the
new issue presented here is the inclusion of the self interactions of
the non-Abelian gauge multiplet. This is interesting in particular
because in order to obtain the vector multiplet propagator a proper
gauge fixing is required. When a generic gauge is applied both the
higher dimensional Lorentz invariance is lost and there one observes a
mixing between the various $\cN=1$ superfields. However, there is a
convenient gauge choice available in which these problems can be avoided 
\cite{Flacke:2003ac,Gates:2005mc}. Using these ingredients we perform
our calculation of brane and bulk gauge operators on 5D and 6D
orbifolds.

Even though our investigation is restricted to one--loop corrections
only, we expect that the results in fact hold to all orders in 
perturbation theory up to infra--red (IR) effects. Both at the fixed
points and in the bulk holomorphicity arguments 
\cite{Shifman:1986zi,Shifman:1991dz,Seiberg:1993vc,Seiberg:1994bp,Weinberg:2000cr}
of $\cN=1$ SUSY field theories in 4D apply. 
Using such arguments the behavior of uncompactified supersymmetric
gauge theories in 5D were investigated  by 
\cite{Seiberg:1996bd,Morrison:1996xf,Intriligator:1997pq} starting
from an anomaly argument by Witten \cite{Witten:1996qb}. 
In the direct perturbative calculation that we will be performing, we
should of course be able to reproduce those results, and so they can
serve as important cross checks.

This paper is organized as follows: In section \ref{sc:class5D} we
give the classical action for a hyper multiplet coupled to a gauge
multiplet in 5D and motivate the gauge fixing we employ. We modify the
action 
such that it is formulated consistently on the orbifold $S^1/\Intr_2$.
We introduce orbifold compatible delta functions which are necessary
for functional differentiation in order to calculate Feynman graphs on
the orbifold. We generalize this concept to 6D and the orbifold
$T^2/\Intr_N$ in section \ref{sc:class6D}. Section \ref{sc:5Dquantum}
describes the quantum calculation of the vector multiplet self-energy
in the 5D case. We present the relevant vertices and calculate the
Feynman graphs. In a detailed example we demonstrate how to compute an
amplitude directly on the orbifold. We take the sum of the 
graphs to obtain the vector multiplet self-energy. We regularize the
result and calculate the bulk and $S^1/\Intr_2$ fixed points 
counterterms. Section \ref{sc:6Dquantum} follows the same logic for
the 6D case and $T^2/\Intr_N$. Evaluating the vector multiplet
self-energy here shows that one has special cancellations at those
fixed points of $T^2/\Intr_N$ that are invariant under a $\Intr_2$
subgroup of $\Intr_N$. In the final section \ref{sc:crosscheck} we
give some illustrating examples in which we relate our results to 4D
on the zero mode level before we conclude. Appendix
\ref{sc:graphs_expr} has all explicit results for the calculated
Feynman graphs, while appendices \ref{sc:fourier} and
\ref{sc:theta_functions} show our conventions for Fourier
transformation and theta functions, respectively. We regularize the
relevant momentum integrals in appendix \ref{Regularization_Integral}.

%%% Local Variables: 
%%% mode: latex
%%% TeX-master: "paper"
%%% End: 

    \section{Hyper and non-Abelian gauge multiplets in five dimensions}             %%% 5D NA %%%
\labl{sc:class5D}

In this section we consider a hyper multiplet charged under a
(non-)Abelian vector multiplet on the orbifold $S^1/\Intr_2$ in 5D. We
begin our discussion with a review of these 5D multiplets using a 4D
superfield language. Next we determine the propagators for these
superfields. For the vector multiplet this requires gauge fixing and
the introduction of ghost multiplets. In the final part of this
section we explain how this theory can be extended to the 5D
orbifold $S^1/\Intr_2$ and introduce orbifold compatible delta
functions that arise from functional differentiation.

  \subsection{Classical hyper and gauge multiplet actions}                    %%% 5D NA %%%
\labl{sc:classAc5D}

We consider the classical theory of a supersymmetric 
5D theory containing a hyper multiplet that is coupled to a gauge multiplet. 
We describe these multiplets in terms of 4D superfields  
\cite{Arkani-Hamed:2001tb,Hebecker:2001ke,Marti:2001iw,Marcus:1983wb}. 
In this language the degrees of freedom of the 5D hyper multiplet are
described by two 4D chiral multiplets $\Phi_+$ and $\Phi_-$. These
fields transform in a given representation (for example the
fundamental or adjoint representation) of the gauge group. The degrees
of freedom of the 5D gauge multiplet are contained in one 4D vector
multiplet $V = V^i T_i$ and one 4D chiral multiplet $S = S^i T_i$
which both transform in the adjoint representation of the gauge
group. Here $T_i$ are the Hermitian generators of the gauge group. The
algebra of these generators  $[T_i, T_j] = f_{ij}{}^k T_k$ defines the
purely imaginary structure coefficients. The Killing metric, denoted
by $\get_{ij}$, will be used to raise and lower adjoint indices, for
example  $f_{ijk} = f_{ij}{}^\ell \get_{\ell k}$.  We denote the trace in the
representation of the chiral multiplets by $\tr$ and the trace in the
adjoint representation by $\tr_{\rep{Ad}}$. The latter is given by 
\(
\tr_{\rep{Ad}}(X Y) =  - f_{ijk} f_{\ell mn} \get^{jm} \get^{kn} X^i Y^\ell,
\) 
where the matrix $X$ and $Y$ are defined in the adjoint: 
$(X)_{jk} = X^i (T_i)_{jk} = X^i f_{ijk}$, etc.

The kinetic action of the hyper multiplet with its coupling to the
gauge multiplet is described by
\equa{
  \label{NA5DSH} \dsp 
  \cS_H = \int d^5x \Big[
\! & \! \dsp 
 \int d^4\theta \left( \bar{\Phi}_+ e^{2V} \Phi_+ + \Phi_- e^{-2V}
   \bar{\Phi}_- \right) + 
\\[2ex] + & \dsp 
  \int d^2\theta \, \Phi_- (\del_5 + \sqrt{2}S) \Phi_+ + \int
    d^2\bgth \, \bar{\Phi}_+ (-\del_5 + \sqrt{2}\bar{S}) \bar{\Phi}_-
    \Big]. \nonumber 
}
Here we have indicated the derivative in the fifth direction by 
$\der_5$. This action is invariant under the supergauge transformations
\be
  \label{NA5DGT1}
  \Phi_+ \rightarrow e^{-2\Lambda} \Phi_+,   \quad  \Phi_-
  \rightarrow \Phi_- e^{2\Lambda},
\qquad 
  S \rightarrow e^{-2\Lambda} \left( S + \frac{1}{\sqrt{2}} \, \del_5 \right) e^{2\Lambda},   \quad  e^{2V} \rightarrow e^{2\bar{\Lambda}}e^{2V}e^{2\Lambda},
\ee
where $\Lambda$ is a chiral superfield and $\bgL$ its conjugate. These
conventions (that lead to various factors of $2$) ensure that the
scalar and fermionic components have charges normalized to unity.

The kinetic action for the 5D gauge multiplet in a 4D superfield
language comprises the standard terms for the 4D gauge field $V$ and
one extra term for the 4D chiral multiplet $S$  
\be
  \label{NA5DSV}
  \cS_V = \frac 1{g^2} \int d^5x \, \tr \left[ \frac{1}{4} \int d^2\gth
    \,\,  W^\alpha W_\alpha  + \frac{1}{4} \int
    d^2\bgth  \,\,\bar{W}_\dga \bar{W}^\dga +
    \frac{1}{4} \int d^4\theta \,\, 
      e^{2V_5}e^{-2V}e^{2V_5}e^{-2V} \right], 
\ee
where we have defined
\be
\label{defV5}
  W_\alpha = -\frac{1}{8} \bar{D}^2 \left( e^{-2V} D_\alpha e^{2V} \right)   \quad\quad  \text{and} \quad\quad  e^{2V_5} = \del_5 e^{2V} - \sqrt{2} e^{2V} S - \sqrt{2} \bar{S} e^{2V}.
\ee
Application of the gauge transformations (\ref{NA5DGT1}) shows that
$W_\ga$ and $e^{2V_5}$ transform covariantly 
\be
  \label{NA5DGT2}
  W_\alpha \rightarrow e^{-2\Lambda} W_\alpha e^{2\Lambda} \quad \quad \text{and} \quad \quad e^{2V_5} \rightarrow e^{2\bar{\Lambda}} e^{2V_5} e^{2\Lambda}
\ee
such that the vector multiplet action is gauge invariant. The
reduction to the Abelian case is trivial, where one finds in
particular that the super field strengths $W_\ga$ and $V_5$ are gauge
invariant. When we compute the renormalization of the vector multiplet
at one loop, we perform a direct computation rather than a background
field method. Therefore, we will be able to recover only the quadratic
part of the vector multiplet action 
\be
  \label{NA5DSV2}
  \cS_{V2} = \frac 1{g^2} \int d^5x \, d^4\gth \, \tr \left[ \frac{1}{8} V D^\ga
    \bD^2 D_\ga V + (\del_5 V)^2 -\sqrt{2} \del_5 V \left( S+\bS
    \right) + \bS S  \right]. 
\ee
This expression is obtained from \eqref{NA5DSV} after some partial
integrations and the absorption of a $-\frac 14 \bD{}^2$ in the
Grassmannian integration measure. There is a mixing between the 4D
vector multiplet $V$ and the chiral multiplet $S$. The presence of
this mixing is not surprising, because $S$ behaves like a Goldstone
superfield since it transforms with a shift under gauge
transformations, see \eqref{NA5DGT1}. From a computational standpoint
this mixing is a nuisance, but luckily, it can be removed by a
suitable choice of gauge fixing, as we discuss below.

This description is clearly not manifestly 5D Lorentz
invariant. Lorentz invariance is recovered after eliminating the
auxiliary fields by their equations of motion. Therefore, this 
description is not an off-shell formulation of the 5D supersymmetric
theories. However, for us the main advantage of this approach is that
perturbation theory is greatly simplified over a component approach
and all kinds of cancellations due to $N=1$ supersymmetry are built
in.

  \subsection{Propagators, gauge fixing and ghosts}                                                  %%% 5D NA %%%
\labl{sc:prop5D}

After this strictly classical discussion of the 5D hyper and vector
multiplets we now turn towards the quantization of the theory using
path integral methods. To this end we need to determine the
propagators of the 4D superfields $\gF_+, \gF_-, V$ and $S$ by
coupling them to the sources $J_+, J_-, J_V$ and $J_S$,
respectively. As usual the interactions can be obtained by functional
differentiation with respect to these sources, after the original
superfields are integrated out using their corresponding quadratic
actions.

By considering the quadratic part of the hyper multiplet action
\eqref{NA5DSH} and using some standard superspace identities,  
we thus obtain 
\be
\labl{propH}
  \cS_\text{H2} = \int d^5x \, \d^4\gth \left( J_+ \bJ_- \right) \frac{-1}{\Box + \del_5^2} \left( \begin{array}{cc} 1 & \del_5 \frac{D^2}{-4\Box} \\ -\del_5 \frac{\bD^2}{-4\Box} & 1 \end{array} \right) \left( \begin{array}{c} \bJ_+ \\ J_- \end{array} \right).
\ee
Hence as for massive chiral multiplets in 4D we have both non-chiral
propagators between $\bJ_\pm$ and $J_\pm$, as well as chiral
propagators between $J_+$ and $J_-$ and their conjugates. 
In figure \ref{fg:Propagators} we depict our drawing conventions
of these chiral propagators: The first propagator in this picture
gives the correlation between the sources $\bJ_\pm$ and $J_\pm$, and
the second one between $J_+$ and $J_-$. Obviously, there is also the
conjugate propagator between $\bJ_+$ and $\bJ_-$. 

\Propagators

For the 5D vector multiplet we need to do more work because of 
gauge invariance. The problem of resulting zero modes can be made
manifest by representing the quadratic action \eqref{NA5DSV2} in the
following matrix form 
\equ{
\label{NA5DSV2x}
\cS_{V2} = \frac{1}{2} \int d^5 x \, d^4 \gth \, \tr
\pmtrx{ V & S & \bS} A \pmtrx{V \\[1ex]  S \\[1ex]  \bS}, 
\qquad 
 A = \pmtrx{ - \Box P_0  - \der_5^2 
& \frac 12\sqrt 2\,P_+ \der_5 
& \frac 12\sqrt 2\, P_- \der_5 
\\[1ex] 
-\frac 12\sqrt 2\,  P_- \der_5 & 0&  \frac 12 P_- 
\\[1ex] 
-\frac 12\sqrt 2\,P_+  \der_5 & \frac 12 P_+ & 0 
},  
}
using the transversal projector
$P_0 = \frac{D^\ga \bD{}^2 D_\ga}{- 8 \Box}$ and its chiral
counterparts $P_+ = \frac{\bar{D}^2D^2}{16\Box}$ and 
$P_- = \frac{D^2\bar{D}^2}{16\Box}$. The operator $A$ has chiral zero
modes corresponding to the gauge directions $X$. Indeed, we see that  
\equ{
X = \gd_\gL \pmtrx{V \\[0.5ex] S \\[0.5ex] \bS} = 
\pmtrx{\gL + \bgL \\[0.5ex] \sqrt 2 \, \der_5\gL \\[0.5ex] \sqrt 2\,
  \der_5\bgL}: 
\qquad 
A \, X = 0. 
}
This shows explicitly that also in five dimensions in order to define the
propagator of the vector multiplet, we need to perform a gauge fixing
to modify the quadratic form $A$ so that it  becomes invertible.

The procedure to determine the gauge fixed action follows the
conventional 4D superfield methods for gauge multiplets, see the textbooks  
\cite{Gates:1983nr,West:1990tg} for example.  As usual we start by
choosing a gauge fixing functional 
\be
  \label{GFCOND5D}
\gTh = \frac{\bar{D}^2}{-4} \left( \sqrt{2}V +
  \frac{1}{\Box}\del_5\bar{S} \right). 
\ee
This gauge fixing functional has been previously considered in refs.\ 
\cite{Flacke:2003ac,Gates:2005mc}. To motivate this choice we observe,
that taking the imaginary part of the restriction 
\be
  \label{NA5DGF1}
  \frac{D^2}{-4} \gTh | = \frac{1}{\sqrt{2}} \left( \Box C + D + \del_5 \varphi - \text{i} \del_M A^M \right)
\ee
reveals that with the gauge fixing functional $\gTh$ 5D
Lorentz invariant gauge fixing like $\del_M A^M = 0$ is incorporated. 
The gauge fixing condition $\gTh = F$, with $F$ an arbitrary chiral
superfield,  is implemented into the path integral via the standard
procedure as the argument of a delta function together with a 
compensating Fadeev-Popov determinant $\gD(\gTh)$. One is free to
include a Gaussian weighting factor 
\(
\exp i \int d^5 x d^4 \gth \, \tr \bF F
\)
and to perform functional integration over $F$. Because of the delta
functions, that implement the gauge fixing, this Gaussian integration is
trivial and results in the gauge fixing action 
\be
  \label{NA5DSGF}
  \cS_\text{gf} = - \int d^5x\;d^4\theta \, \tr \left[ \gTh \bgTh \right].
\ee
Combining this gauge fixing action with \eqref{NA5DSV2x} gives rise to
invertible quadratic operators 
\be
  \label{NA5DSV2b}
  \cS_{V2} + \cS_\text{gf} = \int d^5x\;d^4\theta \,
    \tr \left[ - V \left( \Box + \del_5^2 \right) V + \bS \left( 1 +
      \frac{\del_5^2}{\Box} \right) S \right]. 
\ee
Here we see a further motivation for the gauge fixing functional 
\eqref{GFCOND5D}: The mixing between the $V$ and the $S$
and $\bS$ fields, which was present in (\ref{NA5DSV2x}), has been
removed. Consequently, the propagators for $V$ and $S$ are decoupled 
\be
  \label{NA5DSV2'}
  \cS_\text{V2'} = \int d^5x \, d^4\gth \, \tr \left[ \frac{1}{4} J_V
    \frac{1}{\Box + \del_5^2} J_V + \bJ_S \frac{-1}{\Box + \del_5^2}
    J_S  \right]. 
\ee
This decoupling amounts to a major simplification of the supergraph computations
performed in later sections. Notice that the nonlocal term in
\eqref{NA5DSV2b} has given rise to a perfectly regular propagator for
the $S$ superfield. The propagators are depicted in figure
\ref{fg:Propagators}. As observed above, the superfield $S$ can be
thought of as Goldstone boson superfield, therefore in this sense this
gauge fixing is an application of the supersymmetric 't Hooft $R_\gx$
gauge \cite{Ovrut:1981wa}.

To finish the description of the gauge fixing procedure, we rewrite
the Faddeev-Poppov determinant $\gD(\gTh)$ using ghosts as usual.  In
the supersymmetric setting the ghosts $C$ and $C'$ are anti-commuting
chiral superfields. To determine their action we consider the 
infinitesimal supergauge variations (\ref{NA5DGT1}) of the fields $V$
and $S$ 
\be  \delta_\Lambda V  = L_V \left( \Lambda - \bar{\Lambda} \right) +
\coth \left( L_V \right) L_V \left( \Lambda + \bar{\Lambda}  \right),
\quad \quad \delta_\Lambda S = \sqrt{2} \, \del_5 \Lambda  + 2 \left[
  S, \Lambda \right], 
\ee
that are present in the gauge fixing functional \eqref{GFCOND5D}. 
Here $L_V (X) = [V,X]$ denotes the Lie derivative. These
variations determine the Fadeev-Popov determinant 
\be
  \gD(\gTh) = \int \cD \Lambda \cD \Lambda' \,\, \exp \left( \mbox{$- \frac{\text{i}}{\sqrt{2}} \int d^5x \, \left[ \int d^2\theta \, \Lambda' \delta_\Lambda \Theta + \int d^2\bar{\theta} \, \bar{\Lambda}' \delta_\Lambda \bar{\Theta} \right]$} \right). 
\ee
The inverse of the Fadeev Popov determinant is obtained by 
replacing $\Lambda$ and $\Lambda'$ by the ghosts $C$ and $C'$,
respectively. In this way $\gD(\gTh)^{-1}$ can be written as
exponential of the ghost action  
\be
  \label{NA5DSGH}
\arry{c}{ \dsp 
  \cS_\text{gh} = \frac{1}{\sqrt{2}} \int d^5x\;d^4\theta \,\, \tr \left[ \sqrt{2} \left( C'+\bar{C}' \right) \left( L_V \left( C-\bar{C} \right) + \coth \left( L_V \right) L_V \left( C+\bar{C}  \right) \right) \right.
\\[2ex] \dsp 
  + \left. C' \frac{\del_5}{\Box} \left( \sqrt{2}\del_5\bar{C} - 2
      \left[ \bar{S}, \bar{C} \right] \right) + \bar{C}'
    \frac{\del_5}{\Box} \left( \sqrt{2}\del_5 C + 2 \left[ S, C
      \right] \right) \right].
} 
\ee
From this action the ghost propagators can be read off easily
\be
\labl{propGh}
  \cS_\text{gh2} = \int d^5x \, d^4\gth \, \tr \left[ -\bJ'_C \frac{1}{\Box + \del_5^2} J_C - J'_C \frac{1}{\Box + \del_5^2} \bJ_C  \right].
\ee
Notice that even though the (quadratic) action \eqref{NA5DSGH} appears
to include non-local terms, the ghosts have perfectly normal 5D
propagators. These propagators are given in figure
\ref{fg:Propagators}. Even though there are two types of propagators,
we use only one notation for both of them, because the two propagators
  are the same.

This completes our description of the quantum field theory of hyper
and vector multiplets in 5D. The vertices can be obtained
straightforwardly by expanding the various actions and
will not be given here. In section \ref{sc:5Dquantum} we will only give
those interaction terms that will be relevant for the computations
performed there.

\subsection[The five dimensional orbifold ${S^1/\Intr_2}$] 
{The five dimensional orbifold $\boldsymbol{S^1/\Intr_2}$}                  %%% 5D NA %%%
\labl{sc:S1Z2}

In the discussion so far we have only considered vector and hyper
multiplets in 5D Minkowski space. We now turn to the situation where
the fifth dimension is compactified on the orbifold $S^1/\Intr_2$. As
far as the perturbation theory is concerned we only need to reconsider
the functional differentiation w.r.t.\ the sources $J_\pm$, $J_V$ and
$J_S$. This naturally leads to the definition of orbifold compatible
delta functions.

To describe the orbifold $S^1/\Intr_2$, we begin by defining the circle
$S^1$ by the identifications 
\be
  \label{circle}
  y \sim y+\gL_W, \qquad \gL_W = 2\pi R \, \Intr,
\ee
where $\gL_W$ is the winding mode lattice. The length of the circle
(the ``volume'' of a fundamental region of the lattice $\gL_W$) is
equal to $\Vol_W = 2\pi R$. We denote the delta function on the torus
by 
\(
\gd(y)  = \gd_\Real(y + \gL_W). 
\) 
The momentum in the fifth direction $p^5$ is quantized and
takes values in the Kaluza-Klein lattice such that the 5D
integral is defined as 
\be
  \int \frac{d^5p}{(2\pi)^5} = \int \frac{d^4p}{(2\pi)^4}
  \frac{1}{2\pi R} \sum_{p^5 \in \gL_K}, \qquad \gL_K = \Intr/R. 
\ee
The volume of the Kaluza-Klein lattice is given by 
$\Vol_K = \frac{1}{R}$.

To construct the orbifold $S^1/\Intr_2$, we need to divide out a
$\Intr_2$ point group. We implement the $\Intr_2$ action as a reflection $y
\rightarrow -y.$ This implies that the derivative in the
extra dimension transforms as $\del_5 \rightarrow - \del_5$. The
fundamental domain of the $S^1/\Intr_2$ orbifold is the interval
$[0,\pi R]$. It has two fixed points located at $y=0$ and $y=\pi
R$. The delta function that peaks at these two fixed points is given
by $\gd (2y)$ and can be expanded into 
\be
  \gd (2y) = \frac{1}{2} \left( \gd (y) + \gd (y-\pi R) \right).
\labl{5Ddelta}
\ee
The normalization using the number of fixed points, 2 for
$S^1/\Intr_2$, ensures that the integral of this delta function over
the circle is unity.

To describe the five dimensional hyper multiplet coupled to
the gauge multiplet on this orbifold, the fields have to be orbifold
compatible such that their action is invariant under the orbifold
symmetry. This means that they must transform covariantly under the
orbifold action 
\be
  \label{NA5DORB}
  \gF_+ \rightarrow Z\gF_+, \quad \quad \gF_- \rightarrow - \gF_-Z,
  \qquad V \rightarrow Z V Z, \quad \quad  S \rightarrow - Z S Z. 
\ee
Such orbifold compatible (super)fields and sources can always be
constructed by taking suitable linear combinations of the fields defined 
on the covering space and their $\Intr_2$ reflections.  
Invariance of the action implies that the transformation of the hyper
and vector multiplets are encoded in a single unitary matrix $Z$. 
Because this is a $\Intr_2$ action, the matrix $Z$ fulfills $Z^2 = \Id$.
Hence $Z$ is a real symmetric matrix with the eigenvalues $\pm1$. 
As it is often convenient to make the adjoint indices on $V$ and $S$
explicit, we introduce the matrix $Q^i{}_j$ to write the
transformation rules for the $V$ and $S$ superfields as 
\be
  \label{V5DTRAFO}
V^i \rightarrow Q^i{}_j V^j, 
\qquad 
S^i \rightarrow - Q^i{}_j S^j, 
\qquad 
  Q^i{}_j = \tr [ T^i Z T_j Z ]
\ee
The invariance of the action requires that the matrix $Q$ fulfills 
\be
  \label{Q5DPROP}
  Q^i{}_{i'} \; Q^j{}_{j'} \; \eta_{\;ij} = \eta_{\;i'j'}, \qquad f_{ijk}\,\Q{i}{i'}\,\Q{j}{j'}\,\Q{k}{k'} = f_{i'j'k'} \;,
\ee
such that it is orthogonal with respect to the Killing metric $\eta_{ij}$. 
We infer that all matrix elements $Q^i{}_j$ are real. 
And due to the $\Intr_2$ symmetry we know that $Q^2 = \Id$ and hence 
$Q$ is a real symmetric matrix. In the computation of the one loop
self-energies, see section \ref{sc:5Dquantum}, we will be making
frequent use of the properties of the matrices $Z$ and $Q$.

For such Feynman super graph computations using the path
integral formalism it is important to know the orbifold compatible
delta functions obtained by functional differentiation w.r.t.\ 
orbifold compatible sources: 
\be
  \frac{\gd J_{+2b}}{\gd J_{+1a}} 
= - \frac 14\bD{}^2  \, \tgd_{21}^{(+)}{}^a{}_b,  
\quad 
\frac{\gd J_{-2}{}^b}{\gd J_{-1}{}^a} 
= -\frac 14 \bD{}^2\, \tgd_{21}^{(-)}{}^b{}_a,  
\quad 
\fdu{V}{2}{1}{i}{j} =  \tgd_{21}^{(V)}{}^i{}_j,  
\quad 
 \fdu{S}{2}{1}{i}{j} 
= - \frac 14 \bD{}^2\,  \tgd_{21}^{(S)}{}^i{}_j .
  \label{5D_func_div}
\ee
Because (except for $J_V$) all these sources are chiral, 
the functional differentiation w.r.t.\ them leads to chiral delta
functions in superspace: $- \frac14 \bD{}^2 \gd^4(\gth_2-\gth_1)$.  
For later convenience we have defined the superspace orbifold
compatible delta functions, indicated as $\tgd$, containing full
Grassmann delta functions $\gd^4(\gth_2-\gth_1)$. As a consequence,
the factor $-\frac14 \bD{}^2$ appears explicitly for the chiral
sources in \eqref{5D_func_div}. The $\Intr_2$ properties of
orbifold compatible fields imply that  
\be
  J_+ \rightarrow J_+ Z, \qquad J_- \rightarrow - Z J_-, \qquad J_V{}^i \rightarrow \Q{i}{j}\,J_V{}^j, \qquad J_S{}^i \rightarrow -\Q{i}{j}\,J_S{}^j.
\ee
where we have used the orthogonality of $Q$ in \eqref{Q5DPROP}. From the transformation properties of the sources we infer that the
orbifold compatible delta functions are given by 
\be
\arry{l}{\dsp 
  \tgd_{21}^{(+)}{}^a{}_b = 
  \frac{1}{2} \Big( \gd^{a}{}_{b}\,\gd (y_2-y_1) + Z^a{}_b\,\gd
    (y_2+y_1) \Big) \; \gd^4 (x_2-x_1) \gd^4 (\gth_2-\gth_1),  
\\[2ex] \dsp 
  \tgd_{21}^{(-)}{}^b{}_a 
  = \frac{1}{2} \Big( \gd^{b}{}_{a}\,\gd (y_2-y_1) - Z^b{}_a\,\gd
    (y_2+y_1) \Big) \; \gd^4 (x_2-x_1) \gd^4 (\gth_2-\gth_1), 
\\[2ex] \dsp 
  \tgd_{21}^{(V)}{}^i{}_j = \frac{1}{2}
  \Big( \gd^{i}{}_{j}\,\gd (y_2-y_1) + \Q{i}{j}\,\gd (y_2+y_1) \Big)
  \gd^4 (x_2-x_1) \; \gd^4 (\gth_2-\gth_1), 
\\[2ex] \dsp 
  \tgd_{21}^{(S)}{}^i{}_j = \frac{1}{2}
  \Big( \gd^{i}{}_{j}\,\gd (y_2-y_1) - \Q{i}{j}\,\gd (y_2+y_1) \Big)
  \gd^4 (x_2-x_1) \; \gd^4 (\gth_2-\gth_1). 
}
  \label{5D_delta_functions}
\ee
These delta functions are the key elements of our formalism for
calculating Feynman graphs directly on the orbifold, since they
contain all the geometric information about the orbifold compatible
superfields. Therefore, it is important to develop some of their
properties: All delta functions are symmetric in their spacetime and
gauge indices, while under a reflection of either $y_1$ or $y_2$ the
delta functions transform as
\be
  \label{DELTA5DTRAFO}
  \tgd_{21}^{(\pm)}{}^a{}_b \rightarrow \pm Z^a{}_{a'}\,\tgd_{21}^{(\pm)}{}^{a'}{}_b \;, \quad \quad \tgd_{21}^{(V)}{}^i{}_j \rightarrow Q^i{}_{i'}\,\tgd_{21}^{(V)}{}^{i'}{}_j \;, \quad \quad \tgd_{21}^{(S)}{}^i{}_j \rightarrow -Q^i{}_{i'}\,\tgd_{21}^{(S)}{}^{i'}{}_j \;.
\ee
In calculating amplitudes one often makes use of partial
integration. But as the delta function is a function of two 
coordinates $(x_2,y_2)$ and $(x_1,y_1)$, one sometimes needs to change
the coordinate w.r.t. which a derivative $\der_5$ acts before one can
perform the partial integration.  When this $\del_5$
acts on the delta function, the change of the coordinate may not only bring in a minus
sign as one expects, but may also switch between the types of delta
functions:  
\be
  (\del_5)_2 \, \tgd_{21}^{(\pm)}{}^a{}_b = -(\del_5)_1 \,
  \tgd_{21}^{(\mp)}{}^a{}_b \;, \quad \quad (\del_5)_2 \,
  \tgd_{21}^{(V)}{}^i{}_j = -(\del_5)_1 \, \tgd_{21}^{(S)}{}^i{}_j \;,
  \quad \quad (\del_5)_2 \, \tgd_{21}^{(S)}{}^i{}_j = -(\del_5)_1 \,
  \tgd_{21}^{(V)}{}^i{}_j. 
\ee
With this technology we are ready to perform supergraph computations
on the 5D orbifold $S^1/\Intr_2$ in section \ref{sc:5Dquantum}. But
before that we extend this discussion on the classical level to 6D and
the orbifold $T^2/\Intr_2$.

%%% Local Variables: 
%%% mode: latex
%%% TeX-master: "paper"
%%% End: 

    \section{Hyper and non-Abelian gauge multiplets in
  six dimensions}   %%% 6D NA %%%
\labl{sc:class6D}

In this section we extend our 5D analysis of the previous section to
6D supersymmetric theories on $T^2/\Intr_N$. As this is in principle
straightforward and in order to avoid many repetitions, we only indicate where
we encounter modifications. Most of these modifications have to do
with the question, whether the 5D derivative $\der_5$ has to be
replaced by $\del = \del_5 + i \del_6$ or $\bdel = \del_5 -i\del_6$. 
Here we employ complex coordinates  $z = \frac{1}{2} (x_5 - ix_6)$ and
$\bz = \frac{1}{2}(x_5 + ix_6)$. To make these modifications easy to
trace we use the same structure for this section as was employed in
section \ref{sc:class5D}. Since the properties of the orbifold
$T^2/\Intr_N$ are more complicated than those of $S^1/\Intr_2$, we
describe them more explicitly.

  \subsection{Classical hyper and gauge multiplet actions}                    %%% 6D NA %%%
\labl{sc:classAc6D}

The only terms in the hyper multiplet action \eqref{NA5DSH} that are
changed contain the 5D derivative operator $\der_5$ and take the form: 
\be
  \label{NA6DSH}
  \cS_H \supset \int d^6x \left[ \int d^2\theta \, \Phi_- \del \Phi_+ - \int d^2\bgth \, \bar{\Phi}_+ \bdel \bar{\Phi}_- \right].
\ee
The supergauge transformations are the same as the 5D transformations
(\ref{NA5DGT1}) except for the superfield  $S$, which transforms as
\be
  \label{NA6DGT1}
  S \rightarrow e^{-2\Lambda} \left( S + \frac{1}{\sqrt{2}} \, \del \right) e^{2\Lambda}.
\ee
Notice that in both these expressions the holomorphic derivative $\der$
appears only in those places where chiral superfields are present.

For the vector multiplet the derivative term, i.e.\ the last term in
\eqref{NA5DSV}, has to be modified to 
\be
  \label{NA6DSV}
  \cS_V \supset \int d^6x \int d^4\theta \, \tr \left[ \left(
      -\frac{1}{\sqrt{2}}\bdel + \bS \right) e^{2V} \left(
      \frac{1}{\sqrt{2}}\del + S \right) e^{-2V} + \frac{1}{4}\del
    e^{-2V} \bdel e^{2V} \right].  
\ee
Notice that in the 6D case it is not possible to represent this result
in terms of a single gauge covariant vector superfield like the
superfield $V_5$ defined in \eqref{defV5}. 
In addition to this obvious modification a Wess-Zumino-Witten
term has to be added in order to preserve the supergauge invariance 
\cite{Marcus:1983wb}.

\subsection{Propagators, gauge fixing and ghosts}                                                  %%% 5D NA %%%
\labl{sc:prop6D}

The propagators for the hyper multiplet in 6D,  
\be
  \cS_\text{H2} = \int d^6x \, \d^4\gth \left( J_+ \bJ_- \right)
  \frac{-1}{\Box + \del\bdel} \left( \begin{array}{cc} 1 & \bdel
      \frac{D^2}{-4\Box} \\ -\del \frac{\bD^2}{-4\Box} & 1 \end{array}
  \right) \left( \begin{array}{c} \bJ_+ \\ J_- \end{array} \right), 
\ee
are the direct generalization of the expressions given in \eqref{propH}. 
Only in those places where a single $\der_5$ derivative appears, it is
not automatically obvious if it has to be replaced by $\del$ or
$\bdel$.

For the vector multiplet in 6D the gauge fixing functional
\eqref{GFCOND5D} is generalized to 
\be
  \Theta = \frac{\bar{D}^2}{-4} \left( \sqrt{2}V +
    \frac{1}{\Box}\del\bar{S} \right), 
\ee
and the restriction to the highest component now yields 
\be
  \label{NA6Dgf}
  \frac{D^2}{-4} \Theta | = \frac{1}{\sqrt{2}} \left( \Box C + D 
+ \del_6 A_5 -\del_5 A_6 + \text{i} \del_M A^M \right). 
\ee
Hence the imaginary part gives rise to a 6D Lorentz invariant gauge
fixing for the vector field $A_M$. Following the same computation for
the gauge fixed propagators then gives rise to 
\be
  \cS_\text{V2'} = \int d^6x \, d^4\gth \, \tr \Big[ \frac{1}{4} J_V
  \frac{1}{\Box + \del\bdel} J_V + \bJ_S \frac{-1}{\Box + \del\bdel}
  J_S  \Big]. 
\ee
Since in the 5D propagators \eqref{NA5DSV2'} only $\der_5^2$ are
present, this 6D results is precisely as expected.

Finally, in order to determine the ghost propagators in 6D we have to take
into account the following modifications: The infinitesimal version of the
6D transformation law (\ref{NA6DGT1}) for the superfield $S$ reads
\be
  \delta_\Lambda      S   =  \sqrt{2} \, \del       \Lambda  + 2
  \left[      S,       \Lambda   \right], 
\ee
and requires the last two terms of the ghost action \eqref{NA5DSGH} to
be modified to 
\be
  \label{NA6DSGH}
  \cS_\text{gh} \supset \frac{1}{\sqrt{2}} \int d^6x\;d^4\theta \,\,
  \tr \Big[ C' \frac{\del}{\Box} \left( \sqrt{2}\bdel\bar{C} - 2
      \left[ \bar{S}, \bar{C} \right] \right) + \bar{C}'
    \frac{\bdel}{\Box} \left( \sqrt{2}\del C + 2 \left[ S, C \right]
    \right) \Big]. 
\ee
As for the vector multiplet, this leads to the obvious generalization
of the ghost 5D propagators \eqref{propGh}: 
\be
  \cS_\text{gh2} = \int d^6x \, d^4\gth \, \tr \Big[ -\bJ'_C
  \frac{1}{\Box + \del\bdel} J_C - J'_C \frac{1}{\Box + \del\bdel}
  \bJ_C  \Big]. 
\ee
Thus, we see that the propagators in 6D are to a large extent
simple generalizations of the 5D propagators given in section
\ref{sc:prop5D}. Therefore, we use the same conventions to draw 
the propagators in 6D as given in \ref{fg:Propagators}.

\subsection[The six dimensional orbifold ${T^2/\Intr_N}$]    
{The six dimensional orbifold $\boldsymbol{T^2/\Intr_N}$}                  %%% 6D NA %%%
\labl{sc:T2ZN}

Next we consider the compactification of the 6D multiplets on the
orbifold $T^2/\Intr_N$. Because the torus $T^2$ is compact, the only
possible values for the orbifold order $N$ are $2,3,4,6$, but we will keep
our discussion general here. The torus $T^2$ is defined by the
identifications  
\be
  \label{torus}
  z \sim z+ \gL_W, \qquad \gL_W = 
\gp \Big( R_1 \Intr + R_2 e^{i \gth} \Intr \Big). 
\ee
Here $\gL_W$ denotes the winding mode lattice of the torus with the
volume $\text{Vol}_W =  (2\gp)^2 \, R_1 R_2\, \sin\gth$, where $R_1$
and $R_2$ are the radii of the torus and $\gth$ defines its angle, i.e.\ 
$\gth = \pi/2$ gives the square torus. Inspired by the string
literature, we can introduce the complex structure modulus $U$ and the
\Kh\ modulus $T$ of the torus  
\be
  \gL_W = \pi\;\sqrt{\frac{\Im(T)}{\Im(U)}}\;\left( \Intr + U \Intr \right), \qquad
    U = \frac{R_2}{R_1}e^{i\gth}, \qquad T = i R_1R_2 \sin\gth.
\ee
In terms of these variables the volume of the torus reads 
$\text{Vol}_W =  (2\pi)^2\,\Im(T)$. 
The momenta $p$ and $\bp$ of the torus mode functions $\gps_p(z, \bz) =
e^{i(pz+\bp\bz)}$ are quantized: $p$ lies on the Kaluza-Klein
lattice $\gL_K$ (and $\bp$ on the complex conjugate lattice).  The 6D
momentum integral is defined as
\be
  \int \frac{d^6p}{(2\pi)^6} = \int \frac{d^4p}{(2\pi)^4}
  \frac{1}{(2\pi)^2 \Im(T)} \sum_{p \in \gL_K}, \qquad \gL_K
  = \frac{i}{\sqrt{\Im(T) \Im(U)}} \left( \Intr \bU + \Intr
  \right).  
\ee
The volume of the Kaluza-Klein lattice is given by 
$\text{Vol}_K = {1/\Im(T)}$.

To define the orbifold $T^2/\Intr_N$, we implement the $\Intr_N$ 
action of the orbifold group as $z \rightarrow e^{-i\gvf} z$,  with 
$\gvf = \frac{2\pi}{N}$. Consequently, the holomorphic derivative
$\del$ transforms as $\del \rightarrow e^{i\gvf} \del$. 
The delta function, that peaks at the orbifold fixed points $z_f$, is
given by 
\be
  \gd^2 \left( (1-e^{i\gvf}) z \right) = 
\frac{1}{4 |\sin \frac 12 \gvf|^2} \sum_{f} \gd^2(z-z_{f}), 
\ee
in terms of the torus delta function $\gd^2(z)$.  
The factor $4 |\sin \frac 12 \gvf|^2$ equals the number of fixed points
of the $T^2/\Intr_N$ orbifold.

The hyper and gauge multiplets on the orbifold need to be covariant
w.r.t.\ the $\Intr_N$ orbifold action. Hence, their transformation
behaviour under $z \rightarrow e^{-i\varphi}z$ is found to be
\be
  \label{NA6DORB}
  \gF_+ \rightarrow Z_+\gF_+, \qquad \gF_- \rightarrow \gF_-Z_-, \qquad V \rightarrow Z_+ V \bZ_+, \qquad S \rightarrow \bZ_- S \bZ_+,
\ee 
with the properties $Z_+^N = Z_-^N = 1$, because the transformations
are $\Intr_N$ actions. Invariance of the action requires in addition
that the matrices $Z_+$ and $Z_-$ be unitary are related to
each other via:  $Z_+ Z_- e^{i\varphi} = \Id$. Therefore, we only need
the matrix $Z_+$ in principle, however, it turns out to be convenient
to keep using the notation $Z_\pm$. The transformation rules for the
$V$ and $S$ superfields with the adjoint indices made explicit are
given by  
\be
  \label{V6DTRAFO}
  V^i \rightarrow \Q{i}{j}\,V^j \quad \quad S^i \rightarrow
  e^{+i\varphi} \, \Q{i}{j} \, S^j, \qquad 
  Q^i{}_j = \tr [ T^i Z_+ T_j \bZ_+ ]. 
\ee
This implies that all matrix elements $Q^i{}_j$ are real. Invariance
of the action requires $Q$ to have the properties (\ref{Q5DPROP})
and $Q^N = \Id$ as it defines a $\Intr_N$ action. The reduction to
the $\Intr_2$ orbifold group with $\gvf=\pi$ and  $Z_+ = -Z_- = Z$
is interesting, because then many of the properties of the 5D case,
discussed in subsection \ref{sc:S1Z2}, are recovered.

To obtain the orbifold compatible delta functions for the various
superfields, we write down the transformation behaviour of 
orbifold compatible sources under $z \rightarrow e^{-i\varphi}z$ 
\be
  J_+ \rightarrow J_+ Z_+^{-1} \;, \qquad J_- \rightarrow Z_-^{-1} J_- \;, \qquad J_V{}^i \rightarrow \Q{i}{j}\,J_V{}^j \;, \qquad J_S{}^i \rightarrow e^{-i\varphi}\,\Q{i}{j}\,J_S{}^j,
\ee
where the orthogonality property of $Q$ in \eqref{Q5DPROP} has been
used. This is also reflected in the orbifold compatible delta functions for the
$T^2/\Intr_N$ orbifold 
\be 
  \label{6D_delta_functions}
\arry{l}{\dsp 
  \tgd_{21}^{(+)}{}^a{}_b = 
  \frac{1}{N} \sum_{b=0}^{N-1} \left[ Z_+^b \right]{}^a{}_b\,\delta^2
  (z_2-e^{ib\varphi}z_1) \; \gd^4 (x_2-x_1) \; \gd^4 (\gth_2-\gth_1), 
\\[2ex] \dsp 
  \tgd_{21}^{(-)}{}^b{}_a =
  \frac{1}{N} \sum_{b=0}^{N-1} \left[ Z_-^b
  \right]{}^b{}_a\,\delta^2 (z_2-e^{ib\varphi}z_1) \; \gd^4 (x_2-x_1)
  \; \gd^4 (\gth_2-\gth_1), 
\\[2ex] \dsp 
  \tgd_{21}^{(V)}{}^i{}_j = \frac{1}{N}
  \sum_{b=0}^{N-1} \left[ Q^{-b} \right]{}^i{}_j\,\delta^2
  (z_2-e^{ib\varphi}z_1) \; \gd^4 (x_2-x_1) \; \gd^4 (\gth_2-\gth_1),  
\\[2ex] \dsp 
  \tgd_{21}^{(S)}{}^i{}_j = \frac{1}{N}
  \sum_{b=0}^{N-1} e^{ib\varphi} \left[ Q^{-b}
  \right]{}^i{}_j\,\delta^2 (z_2-e^{ib\varphi}z_1) \; \gd^4 (x_2-x_1)
  \; \gd^4 (\gth_2-\gth_1). 
}
\ee
Under $z_2 \rightarrow e^{-i\varphi}z_2$ these delta functions
transform in the same way as the corresponding sources
\be
  \tgd_{21}^{(\pm)}{}^a{}_b \rightarrow \left[ Z_\pm^{-1}
  \right]^a{}_{a'} \, \tgd_{21}^{(\pm)}{}^{a'}{}_b \;, \qquad
  \tgd_{21}^{(V)}{}^i{}_j \rightarrow \Q{i}{i'} \,
  \tgd_{21}^{(V)}{}^{i'}{}_j \;, \quad \quad \tgd_{21}^{(S)}{}^i{}_j
  \rightarrow e^{-i\varphi} \, \Q{i}{i'} \,
  \tgd_{21}^{(S)}{}^{i'}{}_j, 
\ee
and under $z_1 \rightarrow e^{-i\varphi}z_1$ inversely
\be
  \tgd_{21}^{(\pm)}{}^a{}_b \rightarrow \left[ Z_\pm \right]^a{}_{a'}
  \, \tgd_{21}^{(\pm)}{}^{a'}{}_b \;, \qquad \tgd_{21}^{(V)}{}^i{}_j
  \rightarrow \left[ Q^{-1} \right]{}^i{}_{i'} \,
  \tgd_{21}^{(V)}{}^{i'}{}_j \;, \quad \quad \tgd_{21}^{(S)}{}^i{}_j
  \rightarrow e^{i\gvf} \left[ Q^{-1} \right]{}^i{}_{i'} \,
  \tgd_{21}^{(S)}{}^{i'}{}_j. 
\ee
In contrast to the orbifold compatible delta functions
\eqref{5D_delta_functions} in 5D,  these delta functions are no longer
symmetric in their indices: The exchange of the spacetime labels results in 
\be
  \tgd_{12}^{(\pm)}{}^a{}_b = \tbgd_{21}^{(\pm)}{}^a{}_b \;, \qquad \tgd_{12}^{(V)}{}^i{}_j = \tgd_{21}^{(V)}{}_j{}^i \;, \quad \quad \tgd_{12}^{(S)}{}^i{}_j = \tgd_{21}^{(\bS)}{}_j{}^i,
\ee
because $Z_+$ and $Z_-$ are unitary and $Q$ is orthonormal. Derivatives
with respect to the compactified coordinates always act on the
$\delta^2 (z_2-e^{ib\varphi}z_1)$ factor. Therefore, changing a
spacetime index of such a derivative also changes the type of delta
function as 
\be
  \bdel_2 \, \tgd_{21}^{(\pm)}{}^a{}_b = -\bdel_1\,
  \tbgd_{21}^{(\mp)}{}^a{}_b \;, \qquad \del_2 \,
  \tgd_{21}^{(S)}{}^i{}_j = -\del_1\, \tgd_{21}^{(V)}{}^i{}_j \;,
  \quad \quad \del_2 \, \tgd_{21}^{(V)}{}^i{}_j = -\del_1\,
  \tgd_{21}^{(\bS)}{}^i{}_j. 
\ee
Notice that for the hyper multiplet delta functions also a
complex conjugation is performed.

This completes the discussion of the supersymmetric field theory on the
6D orbifold $T^2/\Intr_N$. We have seen, that even though many
properties are very similar to the ones encountered for the
$S^1/\Intr_2$ orbifold discussed in subsection \eqref{sc:S1Z2}, there
are also some important additional complications in the 6D case.

%%% Local Variables: 
%%% mode: latex
%%% TeX-master: "paper"
%%% End: 

\section{Quantum corrections in the 5D theory}
  \label{sc:5Dquantum}

\VectorVert

This section is concerned with the calculation of the running of the
gauge coupling of the 5D gauge multiplet due to vector and hyper multiplets
on the 5D orbifold $S^1/\Intr_2$. The classical action and the
propagators were given in section \ref{sc:class5D}. Here we first
write down the vertices, after that we evaluate the Feynman graphs that lead
to a correction of the gauge coupling at one loop. The relevant
vertices are obtained by expanding the action: To construct genuine
self energy supergraphs we need three point interactions, and to
generate tadpole (seagull) graphs four point vertices are required. 
Hence it is sufficient for us to expand the action to fourth order in
the fields.

We perform these 
Feynman graph calculations directly on the orbifold with the help of
our orbifold compatible delta functions \eqref{5D_delta_functions}
obtained in section \ref{sc:class5D}. The combination of all these
graphs can be divided into two types: One part of these amplitudes
corresponds to bulk effects that are also present when the 5D theory 
is compactified on the circle $S^1$ rather than on the orbifold
$S^1/\Intr_2$. The other part of the amplitudes is sourced by the
orbifold fixed points. The divergent piece of the bulk amplitudes  
is proportional to the quadratic vector multiplet action
\eqref{NA5DSV2} and therefore leads to the renormalization of the bulk
gauge coupling. The divergent piece of the amplitude sourced by the
fixed gives rise to renormalization of the gauge coupling at the 4D
fixed points. In addition to this, the part of the bulk superfield $S$
that is not projected away at these fixed points receives wave
function renormalization. We calculate the divergences and determine 
the counter terms.

\subsection{Gauge multiplet contributions to the vector multiplet self
  energy} 
\labl{sc:GaugeContr}

In this subsection we compute the one loop vector multiplet self
energy due to the vector multiplet self interactions. Because of the
gauge fixing described in section \ref{sc:prop5D} we encounter the
superfields $V$, $S$ and the ghosts $C$, $C'$ in the loops. After
describing the vertices we list the resulting diagrams.

Performing the expansion to fourth order in the gauge sector
(\ref{NA5DSV}) leads to the following interactions 
\equ{ 
  \gD\cS_\text{V} \supset \int d^5x\,d^4\gth \, \tr \Big[ \frac{1}{4}
    [V,D^\ga V]\bD^2 D_\ga V - \frac{1}{8} [V, D^\ga V] \bD^2 [V,D_\ga
    V]   - \frac{1}{6} [V,[V,D^\ga V]] \bD^2 D_\ga V + 
\non \\[2ex] 
+ \sqrt{2} \del_5
    V [V, \bS-S] -2S[V,\bS] 
+ \frac{1}{3} \del_5 V [V,[V,\del_5 V]] - \frac{2}{3}
    \sqrt{2} \del_5 V [V,[V,S+\bS]] + 2 S [V,[V,\bS]]
  \Big].  
\label{VSbS}
}
To indicate that we display only the terms of the expansion up to
fourth order we use the notation ``$\supset$'' instead of ``$=$''. In
deriving \eqref{VSbS} from \eqref{NA5DSV} we have rewritten the
(anti-)chiral superspace integrals into full superspace integration in
the standard way. 
We use the convention that the derivative operator $\der_5$ only acts
on the field it is immediately adjacent to. The interaction vertices
have been collected in figure \ref{fg:SV_vertices}.

In the ghost sector we obtain the following
interactions from the expansion of (\ref{NA5DSGH})
\be
  \label{VCC}
  \gD\cS_\text{gh} \supset \int d^5x \, d^4\gth \, \tr \left[ (C' + \bC') [V,C-\bC] + \sqrt{2} \frac{\del_5}{\Box} C'[\bS,\bC] - \sqrt{2} \frac{\del_5}{\Box} \bC' [S,C] + \right.
\ee
\be
  \left. + \frac{1}{3} (C'+\bC') [V,[V,C+\bC]] \right]. \nonumber
\ee
These vertices are depicted in figure \ref{fg:SGh_vertices}. 
One might worry about a possible non-locality of the interaction of a $V$ field
with two ghosts $C'$ and $\bC$ in (\ref{VCC}), because the term contains a four
dimensional d'Alembertian operator $\Box$ in the denominator.  
But such terms do not necessarily pose a problem, because physical 
amplitudes may also contain a bunch of supercovariant derivatives,
which give rise to additional $\Box$ operators in the
numerator so that cancellations can take place. 
In our calculation this issue does not arise at all, because it is
impossible to construct one loop corrections to the $\bS S$ self energy
with ghosts in the loop. 
The only graphs that could be constructed would be one loop contributions
that are purely chiral, such that they vanish upon superspace integration. 

\GhostVert

The supergraphs for the gauge corrections due to gauge interactions
consist of the $VV$, $\bS S$ and $V S$ self-energies, depicted in
figures  \ref{fg:VV_from_VSC} and  \ref{fg:SS_from_VS}.  
In the first line of figure \ref{fg:VV_from_VSC} the 
genuine self energy graphs are labeled 
\ref{fg:VV_from_VSC}.A to \ref{fg:VV_from_VSC}.D. 
Because there are two ghost propagator diagrams, 
\ref{fg:VV_from_VSC}.D gives rise to four contributions. 
We will use this notation to refer to these supergraphs throughout
the remainder of this paper. Similarly, we use the notation 
\ref{fg:VV_from_VSC}.E to \ref{fg:VV_from_VSC}.G 
to indicate the tadpole supergraphs in the second line. The
contributions from these tadpole graphs are necessary to cancel non
gauge invariant terms from the total amplitude. The first three graphs
in figures \ref{fg:SS_from_VS} are the $S\bS$ self energy 
diagrams. Finally, figure \ref{fg:SS_from_VS}.D gives the self energy
due to the mixing between $S$ and $V$.

We have calculated all graphs directly on the orbifold. The results
are given in appendix \ref{sc:graphs_expr} as they appear in the more
general calculation in 6D which we discuss in section
\ref{sc:6Dquantum};  the reduction to the 5D case is straightforward. 
Before we turn to
discuss the result of the full amplitude evaluated in the bulk and at
the fixed points, we would like to illustrate the main steps that are
required for the calculation of such supergraphs on orbifolds by
considering one such graph in particular.

    \subsection[Example of supergraph computation on an orbifold:
      ${VV}$ self energy graph due to ${S}$
      superfield] 
{Example of supergraph computation on an orbifold:
      $\boldsymbol{VV}$ self energy graph due to $\boldsymbol{S}$
      superfield} 
    \labl{sc:example}

\VVSelfEnergyVS 
  
To illustrate the self energy computations on orbifolds, we have
chosen the $VV$ self energy contribution due to the chiral superfield
$S$ depicted in figure  \ref{fg:VV_from_VSC}.A. 
As a supergraph it is quite simple and therefore we can focus on
the special issues of computing diagrams on $S^1/\Intr_2$. These
techniques can easily be extended to 6D orbifolds like $T^2/\Intr_N$.

The relevant $SV \bS$ interaction term, given in (\ref{VSbS}), 
is used twice in diagram \ref{fg:VV_from_VSC}.A. To calculate this 
self energy graph the $S$ and $\bS$ superfields are replaced by the 
corresponding sources that act on the exponential of the propagators 
\eqref{NA5DSV2'}. After functional derivations we obtain orbifold
compatible delta functions \eqref{5D_delta_functions} (indicated by
the twiddles), so that the expression for the supergraph
\ref{fg:VV_from_VSC}.A on the orbifold reads 
\equl{
  \text{\ref{fg:VV_from_VSC}.A} 
= 2 \, f_{ijk}f_{\ell mn} \sint{1234} 
 V_1^i V_2^\ell  \;
  \tgd_{31}^{(S)}{}_p{}^j \prop{\eta^{pp'}}{2} \DbD{2}{2}
  \tgd_{32}^{(S)}{}_{p'}{}^n \;  
\times  \\[2ex]   \times  
\; \tgd_{42}^{(S)}{}_q{}^m
  \prop{\eta^{qq'}}{1} \DbD{1}{1} \; \tgd_{41}^{(S)}{}_{q'}{}^k. 
\labl{Adef}
\qquad 
}
Here we have used that in the 5D case there is no distinction between
the orbifold delta functions for $S$ and $\bS$:
 $\tgd^{(\bS)} = \tgd^{(S)}$. 
First we try to replace as many orbifold compatible delta functions by
ordinary delta functions as possible. This is always possible for all
but one delta function. The strategy to replace an orbifold delta
function by an ordinary one is always the same: One expands the
orbifold delta function into a sum and performs a substitution such
that all the summands are equal.

For example, we can replace the first orbifold delta function in the
final factor in the expression \eqref{Adef} for diagram
\ref{fg:VV_from_VSC}.A.  We begin by expanding the first delta
function
\equl{
\text{\ref{fg:VV_from_VSC}.A} 
=  2 \, f_{ijk}f_{\ell mn} \sint{1234} V_1^i V_2^\ell \; 
\tgd_{31}^{(S)}{}_p{}^j
  \prop{\eta^{pp'}}{2} \DbD{2}{2} \;
  \tgd_{32}^{(S)}{}_{p'}{}^n 
\; \times \\[2ex]  \times \; 
\frac 12 \, \Big(\gd_q^m \gd (y_4-y_2) - Q_q{}^m \gd
  (y_4 + y_2)\Big) \gd^4 (x_4-x_2) \gd^4 (\gth_4-\gth_2) \,
  \prop{\eta^{qq'}}{1} \DbD{1}{1} \; 
  \tgd_{41}^{(S)}{}_{q'}{}^k . 
\qquad 
}
We perform the reflection $y_4 \rightarrow -y_4$ to show that
\be
   - Q_q{}^m \int dy_4 \;\; \gd (y_4 + y_2) \; \eta^{qq'} \;
   \tgd_{41}^{(S)}{}_{q'}{}^k \; = \;\; \gd_q^m \int dy_4 \;\; \gd
   (y_4 - y_2) \; \eta^{qq'} \; \tgd_{41}^{(S)}{}_{q'}{}^k, 
\ee
where we have used the transformation properties (\ref{DELTA5DTRAFO}) 
of $\tgd_{41}^{(S)}{}_{q'}{}^k$ and the orthogonality of $Q$ in
(\ref{Q5DPROP}). Here we have not copied the propagators because they
contain $\der_5^2$ which is invariant under this reflection. Substituting 
this back into the original expression, we obtain 
\equl{
\text{\ref{fg:VV_from_VSC}.A} 
= 2 \, f_{ijk}f_{\ell mn}  \sint{1234}  V_1^i V_2^\ell 
 \tgd_{31}^{(S)}{}_p{}^j \prop{\eta^{pp'}}{2} \DbD{2}{2} \;
 \tgd_{32}^{(S)}{}_{p'}{}^n
\;  \times \\[2ex]  \times \; 
 \gd (y_4-y_2) \gd^4 (x_4-x_2) \gd^4(\gth_4-\gth_2)
  \prop{1}{1} \DbD{1}{1} \; \tgd_{41}^{(S)}{}^{mk}. 
\qquad 
}
Hence we have removed the orbifold projection on the first delta
function.

In the same fashion we can remove one of the orbifold delta functions
in the first factor. We choose to make the replacement
\be
  \tgd_{31}^{(S)}{}_{p}{}^j \rightarrow \gd_{p}^j \; \gd (y_3-y_1)
  \gd^4 (x_3-x_1) \gd^4(\gth_3-\gth_1).  
\ee
Now we can integrate over $(x,\gth)_3$ and $(x,\gth)_4$ and are left with
\equl{
\shoveright{
\text{\ref{fg:VV_from_VSC}.A} = 2 \, f_{ijk}f_{\ell mn} \sint{12} V_1^i V_2^\ell \;
  \prop{1}{2} \DbD{2}{2} \; \tgd_{21}^{(S)}{}^{nj} \; \prop{1}{2}
  \DDb{2}{2} \; \tgd_{21}^{(S)}{}^{mk}. 
}
}
We can replace one more orbifold delta function. We choose to expand
the second delta function 
\equl{
\text{\ref{fg:VV_from_VSC}.A} 
=  2 \, f_{ijk}f_{\ell mn} \sint{12} V_1^i V_2^\ell \; 
\prop{1}{2}  \DbD{2}{2} \; \tgd_{21}^{(S)}{}^{nj}
\; \times \\[2ex]   \times \;  
\prop{1}{2}
\DDb{2}{2} 
\frac 12 \, \Big(\eta^{mk} \gd (y_2-y_1) - Q^{mk} \gd (y_2 + y_1)\Big)
\gd^4 (x_2-x_1) \gd^4 (\gth_2-\gth_1).
\qquad 
}
Performing the transformation $y_1 \rightarrow -y_1$ one shows that
\be
  - f_{ijk} \; Q^{mk} \int dy_1 \; V_1^i \; \gd (y_2 + y_1) \;
  \tgd_{21}^{(S)}{}^{nj} = f_{ijk} \; \eta^{mk} \int dy_1 \; V_1^i \;
  \gd (y_2 - y_1) \; \tgd_{21}^{(S)}{}^{nj}. 
\ee
Here we used that both the transformation of $V$ in (\ref{V5DTRAFO})
and of the orbifold compatible delta function in (\ref{DELTA5DTRAFO})
bring in a matrix $Q$. Then we applied the orthogonality property of
$Q$ in (\ref{Q5DPROP}) in order to place the indices of all three
$Q$'s alike. Subsequently, we took advantage of the fact that three
$Q$'s contracted with the structure constants leave the structure
constants invariant as found in (\ref{Q5DPROP}). Thus, we find 
\equl{
\shoveright{
\text{\ref{fg:VV_from_VSC}.A} = 2 \, f_{ijk}f_{\ell mn} \eta^{mk}
\sint{12}  V_1^i V_2^\ell \; 
\prop{1}{2} \DbD{2}{2} \; \gd_{21}^{(S)}{}^{nj} \; 
\prop{1}{2} \DDb{2}{2} \; \gd_{21}. 
}
}
Hence we see that in this diagram we have been able to replace all but
one orbifold compatible delta functions by ordinary delta
functions. The final step in the evaluation of this diagram in the
coordinate space representation is to make the expression local in the
Grassmann variables. Making use of standard identities for the covariant
supersymmetric derivatives, we perform the integration over $\gth_2$ 
\equl{ 
\text{\ref{fg:VV_from_VSC}.A} 
   =  f_{ijk}f_{\ell mn}\get^{mk} \, \int (d^5x)_{12}\;d^4\theta \Big[
   \, - V_1^i \Box P_0 V_2^\ell \,\,\,
   \prop{1}{2} \; \dtwup{5(S)}{21}{n}{j}  \prop{1}{2} \; \gd^{5}_{21} + 
\\[2ex] 
       +  2 \, V_1^i V_2^\ell \,\,\, 
      \prop{\Box_2}{2} \; \dtwup{5(S)}{21}{n}{j}\prop{1}{2} \;
      \gd^{5}_{21}
\Big]. 
\qquad 
} 
Since the expression only contains $\gth_1$, it is local in $\gth_1$
and we simply dropped the subscript ``$1$'' on $\gth$.

The structure of the calculation is the same in 5D and in 6D except for the
fact that the orbifold compatible delta functions involve $N$ summands
instead of two. The result for the 6D counterpart of the example
calculation can be found in \eqref{Afinal} in appendix \ref{sc:graphs_expr}. 
One observes that the reduction of the 6D result to 5D is
straightforward by making use of the fact that $\tgd^{(\bS)} = \tgd^{(S)}$. 
Hence we refer to appendix  \ref{sc:graphs_expr} for the expressions
for the other diagrams in figure \ref{fg:VV_from_VSC}. 

We note that in this example it did not matter which of the
two last orbifold delta functions we replaced, the result is the
same. For some other diagrams that we have computed, however, the
final result depends on which of the last two orbifold delta functions
one replaces. As both possible forms are correct, one can use a linear
combination of the two final expressions to make some cancellations
explicit. This happens in particular if we encounter a $\tgd^{(V)}$
and a $\tgd^{(S)}$. For example we will see in section
\ref{sc:VVSelfEnergyVS} that due to such a cancellation only a bulk
contribution is left over in the supergraphs \ref{fg:SS_from_VS}. For
this reason we have given the expressions for the other diagrams in
appendix \ref{sc:graphs_expr} and in the remainder of the chapter at
the level of two orbifold compatible delta functions.

 \subsection{Vector multiplet renormalization due to self interactions}
\labl{sc:VVSelfEnergyVS}

\SSSelfEnergyVS

Now we turn to discuss the result of the combined amplitude of gauge
multiplet self energy due to self interaction in the bulk and at the
fixed points. This amplitude consists of four parts $\gS_{VV}$,
$\gS_{V S}$, $\gS_{V\bS}$ and $\gS_{\bS S}$, because the
vector multiplet is described by the 4D superfields $V$ and $S$.

The $VV$ self energy arises from the supergraphs
\ref{fg:VV_from_VSC}. Using the results for these graphs given in
appendix \ref{sc:AppGauge}, this self energy is found to be 
\equ{
  \gS_{VV} = f_{ijk} f_{\ell mn} \int ( d^5x )_{12} d^4\gth 
\Big[ -3 \, V_1^i  \Box_2 P_0  V_2^\ell \prop{1}{2} 
\dtwup{5(V)}{21}{m}{j} \prop{1}{2}  \dtwup{5(V)}{21}{n}{k} + 
\labl{gSVV} \\[2ex] 
            +  V_1^i  \Box_2 P_0  V_2^\ell  \prop{1}{2} 
\dtwup{5(\bS)}{21}{m}{j} \prop{1}{2}  \dtwup{5(S)}{21}{n}{k} 
          +2 \, \del_5 V_1^i \del_5 V_2^\ell  \prop{1}{2} 
\dtwup{5(V)}{21}{m}{j} \prop{1}{2}  \dtwup{5(\bS)}{21}{n}{k} 
\Big]. \nonumber
}
This expression still contains two orbifold compatible delta
functions. In the first term in \eqref{gSVV} it does not make a
difference which orbifold delta function we replace by an ordinary
delta function, because they are both of the same type:
$\tgd^{5(V)}$. As the remaining delta function $\tgd^{5(V)}$
contains a sum, see \eqref{5D_delta_functions}, both bulk and
fixed point contributions have the same sign.  For the second
term in \eqref{gSVV}, we conclude that the fixed point contributions
have the opposite sign as compared to the bulk contribution. Only for
the last term it makes a difference which delta function we reduce. To
take possible cancellations into account, we write the amplitude as
half of the sum of both possibilities to reduce one  delta
function. Then the fixed point contribution of the last term in
\eqref{gSVV} vanishes, leaving only a bulk contribution.

For the same reason also the $V S$, $V\bS$ and $S\bS$ self energies, 
given in figure \ref{fg:SS_from_VS}, only have a bulk contribution,
because their two orbifold compatible delta functions expressions are
given by  
\equ{
  \label{gSVS}
  \gS_{V\bS} = - 2\sqrt{2}\, f_{ijk} f_{\ell mn} \int ( d^5x)_{12} 
d^4\gth \, \del_5 V_1^i \bS_2^\ell\,  \prop{1}{2}
\dtwup{5(V)}{21}{m}{j} \, \prop{1}{2} \dtwup{5(\bS)}{21}{n}{k}
}
for $V\bS$, the complex conjugate for $VS$, and
\equ{
\label{gSbSS}
  \gS_{\bS S}  =  2\, f_{ijk} f_{\ell mn} \; \int \left( d^5x \right)_{12}
d^4\gth \, S_1^i \bS_2^\ell \, \prop{1}{2} \dtwup{5(V)}{21}{m}{j}\,
 \prop{1}{2} \dtwup{5(\bS)}{21}{n}{k} 
}
for the $S\bS$ self energy.

By combining these results and expanding the final orbifold compatible
delta functions according to their definitions in
\eqref{5D_delta_functions}, we can identify the bulk and fixed point
contributions. The bulk amplitude is obtained by taking their
summation index $b=0$, and it can be expressed as 
\equa{
  \gS_{\text{bulk}}^{\text{gauge}} = 
f_{ijk} f_{\ell mn} \eta^{mj} \eta^{nk} 
\int ( d^5x )_{12} d^4\gth &\Big[  
- V_1^i \; \Box_2 P_0 \; V_2^\ell + \del_5 V_1^i \del_5 V_2^\ell 
  - \sqrt{2} \del_5 V_1^i (S_2^\ell + \bS_2^\ell\;) + S_1^i \bS_2^\ell
\Big] \, \times
\non \\[2ex] & ~ \times \, 
 \prop{1}{2} \gd (y_2-y_1) \, \prop{1}{2} \gd (y_2-y_1).
  \label{5D_gauge_bulk}
}
The $\Intr_2$ fixed point contributions of \eqref{gSVV}-\eqref{gSbSS}
are simply the $b=1$ terms in the expansion of the last orbifold delta
functions.  As we have explained above, the only non-vanishing
self energy contribution at the fixed points is given by 
\be
  \label{5D_gauge_fp}
  \gS_{\text{fp}}^{\text{gauge}} = 2 f_{ijk} f_{\ell mn}
\eta^{mj} Q^{nk} \int ( d^5x )_{12} d^4\gth \, 
\Big[ - V_1^i \Box_2 P_0 V_2^\ell \Big] \, \prop{1}{2}\gd (y_2-y_1) \,
\prop{1}{2} \; \gd (y_2+y_1). 
\ee
Both the bulk and the fixed point contributions are divergent and
therefore need to be regularized and renormalized. In the next
subsections we perform this task.

\subsubsection{Bulk renormalization} 
\labl{sc:5Dbulkgaugeren}

We now compute the divergent bulk scalar integral corresponding to 
\eqref{5D_gauge_bulk}. Because we need to perform the same analysis
in the 6D situation in section \ref{sc:6Dquantum}, we already employ a
suitable notation which has a straightforward reduction to 5D.

The 5D bulk contribution \eqref{5D_gauge_bulk} has the structure 
\be
  \label{Bulk_Start}
  \cI_D = \int \left(d^Dx\right)_{12} \; A(x_1) B(x_2)
\frac{1}{(\Box+\del\bdel - m^2)_2} \gd_{21} \frac{1}{(\Box+\del\bdel -
m^2)_2} \gd_{21} , 
\ee
with $m$ an infrared regulator mass. 
Here $\gd_{21}$ denotes the delta function on the circle or the torus in
the 5D and 6D case, respectively. (For our application here
in five dimensions one replaces $\del\bdel \rightarrow \del_5^2$ and
uses $z=y$, $\bz=0$.)  We insert a Fourier transformation
(\ref{fourier}) to represent this integral in momentum space as 
\equ{ 
  \cI_D =  \frac{1}{2^2} \int \frac{d^dk}{(2\pi\mu^2)^{d}}
    \sum_{\ell \in \gL_K} \Vol_W A(k,l) B(-k,-l) \, I_D, 
}
where $\text{Vol}_W$ is the volume of the circle or the torus in the 5D
and 6D cases, respectively. The $\gm$ dependence is a result of our
Fourier transformation conventions \ref{sc:fourier}. Here $k$ is the
continuous external momentum in 4D and $n$ the discrete Kaluza-Klein
momentum in the extra dimensions. In order to find the counter terms,
we need to calculate the divergent part of  
\be
  I_{D} = \int \frac{d^dp}{(2\pi)^d} \frac{1}{\Vol_W} \sum_{n \in \gL_K}
\frac{1}{p^2 + |n|^2 + m^2}\frac{1}{(p-k)^2 + |n-l|^2 + m^2}.  
\ee
This has been done in section  \ref{Regularization_Integral}: 
We extend the 4D momentum integral to $d=4 - 2\ge$ dimensions. As
notation we keep $D-d$ to be either 1 or 2, so that also the total
number of dimensions $D$ becomes $\ge$ dependent. 
The divergent part takes the form 
\be
  \label{Bulk_Structure}
  I_{D}\big|_{div} = i \ga_1 + i \ga_2 (k^2 + |l|^2).
\ee
In 5D the second term is not present, i.e.\
\be
  \ga_1 = - \frac{1}{(4\pi)^2} |m|, \qquad \ga_2 = 0, 
\ee
while in 6D we find 
\be
  \ga_1 = \frac{1}{(4\pi)^3} \Big[ 
\Big(\frac{1}{\bge} + \ln \frac{\gm^2}{m^2} \Big) m^2 + m^2 \Big], 
\qquad 
\ga_2 =
\frac{1}{6} \frac{1}{(4\pi)^3} 
\Big(\frac{1}{\bge} + \ln \frac{\gm^2}{m^2} \Big),
\ee
where $\frac 1 \bge = \frac 1 \ge - \gg + \ln 4\pi$. In 6D the second
term in \eqref{Bulk_Structure} is present because $\ga_2 \neq 0$, and
it requires the introduction of a higher dimensional  operator in the
action. Transforming back into position space we obtain the local
terms
\be
  \label{Bulk_End}
  \cI_D\big|_{div} = i \int \d^Dx \; \Big[ \ga_1 \; A(x) B(x) - \ga_2
\; A(x) \left( \Box + \del\bdel \right) B(x) \Big]. 
\ee

Turning back to the 5D case, we find that the local one loop
counterterms which cancel these divergencies read 
\equ{
  \cS_{\text{bulk}}^{\text{gauge}} 
=  \frac{-1}{(4\pi)^2} |m| \,
\int d^5x \; d^4\gth \, \tr_{\rep{Ad}}
\Big[  - V \; \Box P_0 \; V + \del_5 V \del_5 V  
- \sqrt{2} \del_5 V (S + \bS\;) + S \bS \Big].
}

\subsubsection{Fixed points renormalization}
\labl{sc:5Dgaugefp}

Next we discuss the renormalization at the fixed points, starting from
\eqref{5D_gauge_fp}. As in the previous section
\ref{sc:5Dbulkgaugeren}, we perform the discussion such that it can be
applied in both 5D and 6D. The structure of \eqref{5D_gauge_fp} is 
\be
  \label{FP_Start}
  \cJ_D = \int (d^Dx)_{12} \,  A(x_1) B(x_2)\, 
\frac{1}{(\Box+\del\bdel - m^2)_2} 
\gd \Big(z_2 - e^{ik\gvf} z_1 \Big) \, 
\frac{1}{(\Box+\del\bdel - m^2)_2} 
\gd \Big( z_2 -z_1 \Big), 
\ee
with obvious reduction to five dimensions, and where in the delta
function only the compact dimensions have been indicated for
notational simplicity. In momentum space 
\equ{
  \cJ_D =  \frac{1}{2^2} \int \frac{d^dk}{(2\pi\mu)^{2d}}
\sum_{\ell_1,\ell_2} (2\pi)^d
A(k,e^{ik\gvf}\ell_1+\ell_2) B(-k,-\ell_1-\ell_2)
\, J_0.  
}
The divergence is due to the 4D integral
\be
  J_0  = \int \frac{d^dp}{(2\pi)^d} \frac{1}{p^2 +
|\ell_1|^2 + m^2}\frac{1}{(p-k)^2 + |\ell_2|^2 + m^2},  
\ee
which is calculated in (\ref{divergence_4D}). 
One obtains after the transformation into position space 
\be
  \label{FP_End}
  \cJ_D\big|_{div} = \frac{i}{(4\pi)^2} 
\Big( \frac{1}{\bge} + \ln \frac{\gm^2}{m^2} \Big) 
\int d^dx \left( d^2z \right)_{12} \; A(x,z_1) B(x,z_2) \; \gd^2
(z_2-e^{i k\gvf}z_1) \gd^2 (z_2-z_1). 
\ee
This expression is local in the uncompactified 4D directions. In the
compactified dimensions, it is localized on the fixed point, because
of the two delta functions with the two different arguments. 
We apply the result to \eqref{5D_gauge_fp} in order to find the counter
terms that cancel the divergencies on the fixed points
\be
  \cS_{\text{fp}}^{\text{gauge}} 
= \frac{-2}{(4\gp)^2}
\Big(\frac{1}{\bge} + \ln \frac{\gm^2}{m^2} \Big) 
\int d^5x d^4\gth \, \tr_{\rep{Ad}} \Big[
Q V(x,y) \Box P_0 V(x,y) \Big] \, \gd (2y).
\labl{5DZ2fpRenormNA}
\ee

\subsection{Vector multiplet renormalization due to a hyper multiplet}
\labl{sc:HyperContr} 

\HyperVert

The calculation of the hyper multiplet contributions to the vector
multiplet self energy is similar to the computation of the corrections
from the gauge sector. For the hyper multiplet action (\ref{NA5DSH})
the expansion to fourth order gives rise to the interactions 
\be
  \gD\cS_\text{H} \supset \int d^5x \, 
\tr \left[ \int \d^4\gth\, \bgF_\pm ( \pm 2V + 2V^2) \gF_\pm 
+ \int d^2\gth \sqrt{2} \gF_- S \gF_+ + \int d^2\bgth \sqrt{2} \bgF_+
\bS \bgF_- 
\right].
\label{ffV}
\ee
We have depicted the corresponding vertices in figure
\ref{fg:SH_vertices}.

The supergraphs with the hyper multiplet in the loop are depicted in
figure \ref{fg:VV_from_hypers}. The graph
\ref{fg:VV_from_hypers}.A$_\pm$ contains the propagators that 
connect the chiral sources $J_\pm$ with the anti-chiral sources $\bJ_\pm$,
while diagram \ref{fg:VV_from_hypers}.B involves the chiral sources $J_+$ 
and $J_-$.  This diagram also has a Hermitian conjugate partner, 
which we refer to as
$\overline{\mbox{\ref{fg:VV_from_hypers}.B}}$. The tadpole graphs 
are the final diagrams \ref{fg:VV_from_hypers}.C$_\pm$. The $VV$
self-energy takes the form  
\equa{
  \label{Res_5D_Hyper}
  \gS_{VV}   &\, \dsp 
  = \int \left( d^5x \right)_{12} d^4\gth \; \tr \Big[
  -  V_1 \; \prop{1}{2} \; \tgd_{21}^{5(+)} \; \Box_2 P_0 \; V_2
  \prop{1}{2} \; \tbgd_{21}^{5(+)} + 
\\[2ex] &  \dsp 
- V_1 \; \prop{1}{2} \; \tbgd_{21}^{5(-)} \; \Box_2 P_0 \; V_2
  \prop{1}{2} \; \tgd_{21}^{5(-)} 
   + 2 \; \del_5 V_1 \; \prop{1}{2} \; \tgd_{21}^{5(+)} \; \del_5 V_2
   \prop{1}{2} \; \tgd_{21}^{5(-)} \; \Big]. 
\non 
}
The $V\bS$ and $S\bS$ self-energies are given by 
\equa{
  \label{Res_5D_Hyper_VS}
  \gS_{V\bS} = &\,  -2 \sqrt{2} \int \left( d^5x \right)_{12} d^4\gth
  \; \tr \Big[ \; \del_5 V_1 \; \prop{1}{2} \; \tgd_{21}^{5(+)} \;
  \bS_2 \prop{1}{2} \; \tgd_{21}^{5(-)} \; \Big],
\\[2ex]
  \label{Res_5D_Hyper_SbS}
  \gS_{S\bS} =&\, 2 \int \left( d^5x \right)_{12} d^4\gth \; \tr \Big[
  \; S_1 \; \prop{1}{2} \; \tgd_{21}^{5(+)} \; \bS_2 \prop{1}{2} \;
  \tgd_{21}^{5(-)} \; \Big], 
}
where the $VS$ term is just the complex conjugate of the result for
$V\bS$. The corresponding diagrams are given in figures
\ref{fg:SS_from_Hypers}. Here it is interesting that it does not
make a difference which last delta function is removed. This is in
contrast to the self-energy results from the gauge sector where one had
to be careful not to miss important cancellations.

\VVSelfEnergyHypers

The bulk amplitude is found by replacing one more orbifold delta
function, expanding the remainind orbifold delta function and taking
the $b=0$ contribution
\equ{
  \gS^{\text{hyper}}_{\text{bulk}} 
= \int \left( d^5x \right)_{12} d^4\gth \; \tr \big[ -
  V_1 \; \Box_2 P_0 \; V_2 + \del_5 V_1 \del_5 V_2 - \sqrt{2} \del_5
  V_1 (S_2 + \bS_2\;) + S_1 \bS_2 \big] 
\; \times \non \\[2ex]    \times \; 
\prop{1}{2} \; \gd (y_2-y_1) \; \prop{1}{2} \; \gd (y_2-y_1).
  \label{5D_hyper_bulk}
}
By adding the contributions with one of the two orbifold delta
functions removed we find that at the fixed points 
\equ{
 \gS^{\text{hyper}}_{\text{fp}} 
= \frac 12 \int \left( d^5x \right)_{12} d^4\gth \; \tr \Big\{
[\del_5 V_1, Z] \del_5 V_2 - \sqrt{2} [\del_5 V_1, Z] (\bS_2 - S_2\;) 
+ [S_1, Z] \bS_2 \Big\}  
\; \times \non \\[2ex] \times \; 
\prop{1}{2} \; \gd (y_2-y_1) \; \prop{1}{2} \; \gd (y_2+y_1).
  \label{5D_hyper_fp}
 }
This shows that in the case when $Z$ is proportional to the identity and in the
Abelian case the amplitude vanishes at the fixed points.

  \subsubsection{Bulk renormalization}
\labl{sc:5Dbulkhyper}

As in subsection \ref{sc:5Dbulkgaugeren} we can extract the divergent
parts and determine the counter terms, which leads to
\be
  \cS^{\text{hyper}}_{\text{bulk}} = 
\frac{1}{(4\pi)^2} |m| \int d^5x d^4\gth \; \tr
\Big[ - V \; \Box P_0 \; V + \del_5 V \del_5 V - \sqrt{2} \del_5 V (S
+ \bS\;) + S \bS \Big]
\ee
for the correction due to the hyper multiplet.

  \subsubsection{Fixed Points renormalization}
\labl{sc:5Dhyper}

At the fixed points we can write the counter terms as 
\be
  \cS^{\text{hyper}}_{\text{fp}} = 
\frac{-1}{2(4\pi)^2} \Big( \frac{1}{\bge} +  \ln \frac{\gm^2}{m^2}  \Big) 
\int d^5x d^4\gth \; \tr Z 
\big[ \bS - \sqrt 2 \der_5 V, S - \sqrt 2 \der_5 V \big]  
\; \gd(2 y), 
\labl{5DZ2fpRenormH}
\ee
after we have extracted the 4D divergent parts. As we saw at the level
of the amplitude, in the Abelian case the hyper multiplet does not
induce a correction at the fixed points. The  $(\der_5 V)^2$ parts of
this expression have been obtained before, see  \cite{Cheng:2002iz}.

Moreover, note that this expression is not gauge invariant. The
non-linear extension 
\equ{
  \cS^{\text{hyper}}_{\text{fp\,n.l.\,ext}} = 
\frac{-1}{2(4\pi)^2} \Big( \frac{1}{\bge} +  \ln \frac{\gm^2}{m^2}  \Big) 
\int d^5x d^4\gth \; \tr Z 
\Big[ \Big(\bS - \frac 1{\sqrt 2} \der_5\Big) e^{2V}, \Big(S + \frac 1{\sqrt 2} \der_5\Big) e^{-2V} \big]  
\; \gd(2 y), 
\labl{5DZ2fpRenormHnlext}
}
is gauge invariant w.r.t.\ the zero mode supergauge group, which is
defined by $\der_5 \gL = \der_5 \bgL=0$ and $[Z, \gL] = [Z,\bgL] =0$. 
The second condition is a consequence of the orbifold projection at
the $\Intr_2$ fixed points. However, for the full supergauge group at
the fixed points $\der_5 \gL$ and $\der_5 \bgL$ do not necessarily
vanish and do not commute with $Z$. Consequently this expression, as it
stands, is not gauge invariant under the full bulk gauge
transformations. As we will speculate below \eqref{6DgaugeNonZ2} this
might be cured by a Wess-Zumino-Witten-like term.

\SSSelfEnergyHypers

%%% Local Variables: 
%%% mode: latex
%%% TeX-master: "paper"
%%% End: 

  \section{Quantum corrections in the 6D theory}
  \label{sc:6Dquantum}

The approach to calculate the gauge coupling running on the 6D orbifold
$T^2/\Intr_N$ parallels the procedure described in 5D. Therefore,  it will
suffice for us to indicate the points that deviate from our treatment
in the preceeding chapter.

  \subsection{Gauge multiplet contributions to the vector multiplet
    self energy}                       %%% 6D NA %%% 

The following gauge field self interactions (\ref{VSbS}) change,
because they contain the 5D derivative operator $\del_5$ 
\be
  \gD\cS_\text{V} \supset \int d^6x\,d^4\gth \, \tr \left[ \sqrt{2}
    \del V [V, \bS] - \sqrt{2} \bdel V [V, S] + \right. 
\ee
\be
  \left. + \frac{1}{3} \bdel V [V,[V,\del V]] - \frac{2}{3} \sqrt{2}
    \del V [V,[V,\bS]] - \frac{2}{3} \sqrt{2} \bdel V [V,[V,S]]
  \right] \nonumber 
\ee
and the interaction in the ghost sector (\ref{VCC}) changes as
\be
  \gD\cS_\text{gh} \supset \int d^6x \, d^4\gth \, \tr \left[ \sqrt{2}
    \frac{\del}{\Box} C'[\bS,\bC] - \sqrt{2} \frac{\bdel}{\Box} \bC'
    [S,C] \right]. 
\ee
We already mentioned above that in 6D gauge invariance requires the
presence of an additional WZW term for the gauge multiplet $V$
\cite{Marcus:1983wb}. This term leads in principle to a three point
gauge field self interaction. However, it turns out that all graphs
that can be constructed with this additional interaction add up to
zero because of the symmetry of the structure constants. Thus, for our
calculation, in 6D we are left with the same set of relevant graphs as
in the 5D situation.

  \subsection{Vector multiplet renormalization due to self interactions}

The resulting expressions for the amplitudes are consequently also very
similar to the ones given in the 5D case on the orbifold $S^1/\Intr_2$
which we discussed in section \ref{sc:5Dquantum}. There are of course the
obvious modifications of the dimensionality of the integration measure
and $\der_5^2 \ra \der \bder$. In particular, the effects of
the vector multiplet self interaction given in section
\ref{sc:VVSelfEnergyVS} are modified as follows: The $\gS_{VV}$
self energy is the same as in \eqref{gSVV} except for the term that
involves $\der_5$ derivatives w.r.t.\ the fifth dimension. That term is
modified to 
\equ{
  \gS_{VV} \supset f_{ijk} f_{\ell mn} \int ( d^6x )_{12} \; d^4\gth 
\Big[ 2 \, \del V_1^i \bdel V_2^\ell  \propsix{1}{2} 
\dtwup{6(V)}{21}{m}{j} \propsix{1}{2}  \dtwup{6(\bS)}{21}{n}{k} 
\Big].
}
Also the mixing between $V$ and $S$ given by the amplitude 
\eqref{gSVS} involves a derivative. In the 6D case it reads 
\equ{
  \gS_{V\bS} \supset - 2\sqrt{2} \, f_{ijk} f_{\ell mn} \int ( d^6x)_{12} \; 
d^4\gth \, \del V_1^i \bS_2^\ell\,  \propsix{1}{2}
\dtwup{5(V)}{21}{m}{j} \, \propsix{1}{2} \dtwup{5(\bS)}{21}{n}{k}. 
}
The amplitude $\gS_{\bS S}$ does not involve any single $\der_5$, so
that its generalization to 6D is obvious.

  \subsubsection{Bulk renormalization}

Taking these modifications into account we find the following
expression for the counter terms in the bulk
\equ{
  \cS^{\text{gauge}}_{\text{bulk}} =
\frac{2m^2}{(4\pi)^3N}
\Big( \frac{1}{\bge} + \ln \frac{\gm^2}{m^2} + 1 \Big)
\int d^6x d^4\gth \, \tr_{\rep{Ad}}
\Big[ - V \; \Box P_0 \; V + \del V \bdel V - \sqrt{2}
(\del V \bS + \bdel V S\;) + S \bS \Big] + \non \\[2ex]
  - \frac{1}{3\;(4\pi)^3N}
\Big( \frac{1}{\bge} + \ln \frac{\gm^2}{m^2} \Big) \int d^6x d^4\gth \,
\tr_{\rep{Ad}}
\Big[ - V \; \Box P_0 (\Box+\del\bdel) \; V + \del V
(\Box+\del\bdel) \bdel V + \non \\[2ex] 
  - \sqrt{2} \; \del V (\Box+\del\bdel) \bS - \sqrt{2} \; \bdel
  V (\Box+\del\bdel) S + S (\Box+\del\bdel) \bS \Big].  
\labl{6DZNfpRenormNA}
}
We note that the expression in the second line is the quadratic
approximation of the kinetic action of the vector multiplet, see 
\eqref{NA5DSV2}. The remaining part of this expression corresponds to
the renormalization of the quadratic approximation of the higher
derivative term.

We have also encountered these effects in the Abelian case in 5D and
6D, which we studied in \cite{GrootNibbelink:2005vi}. By gauge
invariance we can infer some additional effects. As we reminded the
reader below \eqref{NA6DSV}, the action is not gauge invariant unless
also a Wess-Zumino-Witten term is added \cite{Marcus:1983wb}. 
Therefore, to preserve gauge invariance, also this Wess-Zumino-Witten
term has to be renormalized. Moreover, because also a higher
derivative operator is generated, also a higher derivative analogue of
the Wess-Zumino-Witten term must exist and renormalize. We have not
performed an explicit calculation to confirm the
renormalization. However, we can say that the Wess-Zumino-Witten term
and its higher derivative counterpart will have to renormalize with
the same multiplicative coefficients as the corresponding terms in the
quadratic part of the action in order for the theory to be gauge
invariant at the one-loop level.

  \subsubsection[$\boldsymbol{\Intr_2}$ fixed point renormalization]
{$\boldsymbol{\Intr_2}$ fixed point renormalization}

Even ordered orbifolds have fixed points which are invariant under a
$\Intr_2$ symmetry, where special cancellations take place that are
not present at the other fixed points. For these $\Intr_2$ fixed
points we find the counter term
\equ{
  \cS^{\text{gauge}}_{\text{$\Intr_2$}} =
\frac{-4}{(4\pi)^2N}
\Big( \frac{1}{\bge} + \ln \frac{\gm^2}{m^2} \Big)
\int d^6x \; d^4\gth \, \tr_{\rep{Ad}} \Big[ - Q^{N/2} \; V \; \Box P_0 \; V \Big] \; \gd^2 \big(2z\big).
}

  \subsubsection[Non-$\Intr_2$ fixed point renormalization]
{Non-$\boldsymbol{\Intr_2}$ fixed point renormalization}

At the non-$\Intr_2$ fixed points we obtain instead the result 
\equ{
  \cS^{\text{gauge}}_{\text{non-$\Intr_2$}} =
\frac{1}{(4\pi)^2N} \Big( \frac{1}{\bge}
  + \ln \frac{\gm^2}{m^2} \Big) \sum_{b=1}^{[N/2]_*} \int d^6x \;
  d^4\gth \, \tr_{\rep{Ad}} \Big\{ \non \\[2ex]
\big(  6 \cos bH -\cos b(H+\gvf) -\cos b(H-\gvf) \big)
   \big( - V \Box P_0 V \big) + \non \\[2ex]
+ 2 \cos b(H+\gvf)
  \big( \del V \bdel V -\sqrt 2 \der V \bS 
-\sqrt 2 S \bder V  + S\bS
  \big) + 
\non \\[2ex]
+ 2 \cos bH
  \big( \del V \bdel V -\sqrt 2 \der V \bS 
-\sqrt 2 S \bder V  + S \bS
  \big) 
\Big\} \gd^2 \big((1-e^{ib\gvf}) z \big), 
\labl{6DgaugeNonZ2}
}
where we introduced the hermitean matrix $H$ via $Q=e^{iH}$. 
To arrive at this expression we have used that the matrices $\cos bH$,
etc.\ are symmetric, which is a consequence of the fact that $Q$ is 
orthogonal. The symbol $[N/2]_*$ is defined as $[N/2]_* = \frac
{N-2}{2}$ for $N$ even and $[N/2]_* = \frac {N-1}{2}$ for $N$ odd.  
Because we have only computed a two point function, this expression
for the one loop counterterm is clearly not gauge invariant. Inspired
by the expression \eqref{NA6DSV}, we expect that the non-linear form
of \eqref{6DgaugeNonZ2} is given by 
\equ{
  \cS^{\text{gauge}}_{\text{non-$\Intr_2$\,n.l.\,ext}} =
\frac{1}{(4\pi)^2N} \Big( \frac{1}{\bge}
  + \ln \frac{\gm^2}{m^2} \Big) \sum_{b=1}^{[N/2]_*} \int d^6x \;
  \tr_{\rep{Ad}}\Big\{ \non \\[2ex]
\big(  6 \cos bH -\cos b(H+\gvf) -\cos b(H-\gvf)
 \big) \Big( 
\frac 14 \int d^2 \gth W^\ga W_\ga + \frac 14 \int d^2\bgth  
\bW_\dga \bW^\dga 
 \Big)
+ \non \\[2ex]
+ 2 \cos b(H+\gvf) 
\Big[ 
\Big( \frac{1}{\sqrt{2}}\del + S \Big) e^{-2V} 
\Big( -\frac{1}{\sqrt{2}}\bdel + \bS \Big) e^{2V} 
+ \frac{1}{4} \del e^{-2V}\bdel e^{2V}
\Big] + 
\non \\[2ex]
+ 2 \big( \cos bH \big)
\Big[ 
\Big( -\frac{1}{\sqrt{2}}\bdel + \bS \Big) e^{2V} 
\Big( \frac{1}{\sqrt{2}}\del + S \Big) e^{-2V} 
+ \frac{1}{4} \bdel e^{2V}\del e^{-2V}
\Big]
\Big\} \gd^2 \big((1-e^{ib\gvf}) z \big). 
} 
This is expression is gauge invariant under the zero mode supergauge
group, as defined below \eqref{5DZ2fpRenormHnlext}. However, as was
discussed there also this term is not gauge invariant under the full
supergauge transformations. This is not surprising when one takes into
account that \eqref{NA6DSV} is also not gauge invariant by itself:
One needs to add a Wess-Zumino-Witten term to make the theory gauge
invariant. Therefore we expect that also the expression above can be
made gauge invariant by adding a suitable extension of a
Wess-Zumino-Witten interaction.

\subsection{Vector multiplet renormalization due to a hyper multiplet}

The expansion of the hyper multiplet action to fourth order in the
gauge coupling \eqref{ffV} remains valid in 6D. The self energies
$\gS_{VV}$, $\gS_{V S}$, $\gS_{V\bS}$ and $\gS_{\bS S}$ are the same
as in 5D except for the changes of the derivative operator in the
quadratic part of the vector multiplet action which leads to the
following replacements in \eqref{Res_5D_Hyper} and  \eqref{Res_5D_Hyper_VS}
\equa{
  \label{Res_6D_Hyper}
  \gS_{VV}   &\, \dsp 
  \supset \int \left( d^6x \right)_{12} d^4\gth \; \tr \Big[
  2 \; \del V_1 \; \propsix{1}{2} \; \tgd_{21}^{6(+)} \; \bdel V_2
   \propsix{1}{2} \; \tgd_{21}^{6(-)} \; \Big],
}
\equa{
  \label{Res_6D_Hyper_VS}
  \gS_{V\bS} = &\,  -2 \sqrt{2} \int \left( d^6x \right)_{12} d^4\gth
  \; \tr \Big[ \; \del V_1 \; \propsix{1}{2} \; \tgd_{21}^{6(+)} \;
  \bS_2 \propsix{1}{2} \; \tgd_{21}^{6(-)} \; \Big],
}
while (\ref{Res_5D_Hyper_SbS}) stays the same in 6D. After the
  reduction of one more orbifold projection the bulk
  amplitude for $b=0$ and the amplitude at the $\Intr_2$ fixed
  points of an even ordered orbifold for $b=N/2$ are calculated
  straightforwardly as in 5D.

  \subsubsection{Bulk renormalization}

We extract the divergence and determine the local bulk counter term in 6D
\equa{
  \cS^{\text{hyper}}_{\text{bulk}} = &
\frac{-2m^2}{(4\pi)^3N} \Big( \frac{1}{\bge} + \ln \frac{\gm^2}{m^2} + 1 \Big) \int d^6x d^4\gth \; \tr
\Big[ - V \; \Box P_0 \; V + \del V \bdel V - \sqrt{2} (\del V \bS 
+ \bdel V S\;) + S \bS \Big] + \non \\[2ex]
  &~ + \frac{1}{3\;(4\pi)^3N} \Big( \frac{1}{\bge} + \ln \frac{\gm^2}{m^2} \Big) \int d^6x d^4\gth \; \tr
\Big[ - V \; \Box P_0 (\Box+\del\bdel) \; V + \del V (\Box+\del\bdel) \bdel V + \non \\[2ex]
  &~ - \sqrt{2} \; \del V (\Box+\del\bdel) \bS - \sqrt{2} \; \bdel V (\Box+\del\bdel) S + S (\Box+\del\bdel) \bS \Big]. 
}
In the second and third lines we find the higher dimensional operator
which we already alluded to above equation \eqref{Bulk_End}.

  \subsubsection{$\Intr_2$ fixed point renormalization}

The following counter term is located at the $\Intr_2$ fixed points of
an even ordered orbifold 
\equ{
  \cS^{\text{hyper}}_{\text{$\Intr_2$}} =
\frac{-1}{(4\pi)^2N} \Big( \frac{1}{\bge} + \ln  \frac{\gm^2}{m^2} \Big)
\int d^6x \; d^4\gth \; \tr Z_+^{N/2} \Big( 
\big[ \bS -\sqrt{2} \bdel V, S-\sqrt{2} \del V \big] - \big[ \bdel V, \del V \big] \Big) \gd^2 \big(2z\big).
}
We note that we recover the factors of the quadratic contribution of
\eqref{NA6DSV} enclosed in commutators. As discussed below
\eqref{6DgaugeNonZ2} 
it is possible to construct a non-linear extension of this term which
is invariant under the zero mode gauge group, but such an expression
is not gauge invariant under the full bulk gauge transformations.

\subsubsection{Non-$\Intr_2$ fixed point renormalization}

The counter term at the non-$\Intr_2$ fixed points involves the delta function $\gd^2 \big((1-e^{ib\gvf}) z
\big)$ which is symmetric under a reflection of $b$. By summing the
contributions to $b$ and $-b$ explicitly and introducing the algebra element $A_+$
that corresponds to the unitary matrix $Z_+$ via $Z_+ \equiv e^{iA_+}$
the local counter term can be written as
\equa{
  \cS^{\text{hyper}}_{\text{non-$\Intr_2$}} & =
\frac{-2}{(4\pi)^2N} \Big( \frac{1}{\bge} + \ln \frac{\gm^2}{m^2}
\Big) \sum_{b=1}^{[N/2]_*}  
\int d^6x \; d^4\gth \; \tr \Big[ - \big( \cos b(A_++\gvf) + \cos bA_+
\big) V \Box P_o V + 
\non \\[2ex] & ~+ \cos \; b(A_++\gvf) \big( \del V \bdel V - \sqrt{2} \;
\del V \bS - \sqrt{2} \; S \bdel V  + S\bS \big) 
\non \\[2ex] & ~ + \cos \; (bA_+) \big( \bdel V \del V  - \sqrt{2} \; \bS
\del V - \sqrt{2} \; \bdel V S + \bS S \big) \Big] \gd^2 \big( 
(1-e^{ib\gvf}) z \big).
\labl{6DZNfpRenormH}
}
By formally replacing the matrix $A_+$ by a
scalar $a_+$ and  the trace $\tr$ by the square of the charge $q^2$
one obtains  the abelian result found previously in
\cite{GrootNibbelink:2005vi}. Here we can make the same comments about
non-linear extensions and gauge invariance as below \eqref{6DgaugeNonZ2}.

%%% Local Variables: 
%%% mode: latex
%%% TeX-master: "paper"
%%% End: 

  \section{Some examples as cross checks}
\labl{sc:crosscheck}

In this section we would like to give a few illustrative applications
of the general formulae for the gauge couplings discussed in this
paper. In addition, we use these examples to perform some simple cross
checks of our general results. These checks are inspired by the
discussions in \cite{Hebecker:2002vm} to determine the fixed
point beta functions using zero mode results on orbifolds and their 
covering spaces.

\subsection[Hyper multiplet on ${S^1/\Intr_2}$]{Hyper multiplet on $\boldsymbol{S^1/\Intr_2}$}
\labl{sc:hyperZ2}

We consider an $\SU{N}$ supersymmetric gauge theory in 5D 
on $S^1/\Intr_2$. Besides the vector multiplet we take a single hyper
multiplet in the fundamental representation of $\SU{N}$. The matrix
$Z$ that defines the orbifold action \eqref{NA5DORB} can be chosen
to be diagonal
\equ{
Z = \pmtrx{\Id_{N_0} & \\[1ex] & - \Id_{N_1}}~,
\labl{ZonZ2}
}
where $N_0+N_1 = N$. At the fixed points and in the effective 4D
theory the gauge symmetry is then broken to
\equ{
\SU{N} \ra \SU{N_0} \times \SU{N_1} \times \U{1}. 
}
To compute the bulk and fixed point gauge
coupling renormalization we can use the results given in section
\ref{sc:HyperContr}. In particular we found in subsection \ref{sc:5Dhyper}
that there is no gauge coupling renormalization at the fixed
points.

We now present a way to check this result by comparing it with the
results for the zero modes on the circle $S^1$ and the orbifold
$S^1/\Intr_2$. First of all the renormalization of the gauge couplings
can be directly computed by considering the zero mode spectrum. For
the hyper multiplet we can write 
\equ{
\gF_+ = \pmtrx{ \gF_{+0} \\[1ex] \gF_{+1} }, 
\qquad 
\gF_- = \pmtrx{ \gF_{-0} & \gF_{-1} }. 
}
These components transform under the orbifold action as 
\equ{
\gF_{+I} \ra (-)^I\, \gF_{+I}, 
\qquad 
\gF_{-I} \ra (-)^{I+1}\, \gF_{-I},
} 
with $I = 0,1$, hence the zero modes $\gF_{+0}$ and $\gF_{-1}$ form 
$\rep{N}_0$ and $\crep{N}_1$ representations of 
the zero mode group $\SU{N_0} \times \SU{N_1}$, respectively. Since
these representations are the fundamental and anti-fundamental of these
gauge groups, we obtain by using the standard beta-functions for super
Yang-Mills theories
\equ{
\frac 1{g_0^2(\gm)} = \frac 1{g_0^2} - \frac 1{16\gp^2} \ln \gm^2,
\qquad
\frac 1{g_1^2(\gm)} = \frac 1{g_1^2} - \frac 1{16\gp^2} \ln \gm^2. 
\labl{5DzerocouplingsH}
}
Here we denote with $g^2_I$ the coupling at scale $\gm = 1$, and
$g_I^2(\gm)$ the coupling at scale $\gm$. 
In our discussion we disregard the gauge coupling running of the
$\U{1}$ factor, as it does not play a significant role in the cross
check we consider below.

We can obtain these results from our local results: By integrating
over the orbifold we obtain a relation between the 4D zero mode
gauge couplings $g_0^2$, $g_1^2$, the 5D bulk gauge coupling $g_5^2$
and the local fixed point gauge couplings $g_{f.p.}^2$: 
\equ{
\frac 1{g_0^2(\gm)} = \frac 12 \frac{2\gp R}{g_5^2}(\gm) 
+ \frac 14 \sum_{f.p.} \frac 1{g_{f.p.0}^2(\gm)}, 
\qquad 
\frac 1{g_1^2(\gm)} = \frac 12 \frac{2\gp R}{g_5^2}(\gm) 
+ \frac 14 \sum_{f.p.} \frac 1{g_{f.p.1}^2(\gm)}. 
\labl{5D4DcouplingsH}
}
Here we have included the appropriate factors of $1/2$, because our
results are stated on the covering circle of size $2\gp R$ rather than
on the fundamental domain of the orbifold. In addition, the definition
of the orbifold delta function \eqref{5Ddelta} contains a factor of
$1/2$. Since we have not included Wilson-lines in our analysis, the
quantum corrections at the two fixed points are the same, so that the
sum simply gives a factor of two. In fact, we conclude from
\eqref{5DZ2fpRenormH} that there is no renormalization at the
orbifold fixed points, i.e.\ $g^2_{f.p.I}(\gm) = g^2_{f.p.I}$ is
constant. As our notation is indicating $2\pi R/g_5^2$ can be
interpreted as a 4D coupling that runs with $\gm$. To find this
dependence, we note that in the bulk there is no difference between
the theory on the orbifold $S^1/\Intr_2$ and on the circle $S^1$. On
the circle we find at the zero mode level one full hyper multiplet in
the fundamental of $\SU{N}$, hence
\equ{
\frac {2\pi R}{g_5^2}(\gm) = \frac{2\pi R}{g_5^2} 
- \frac 2{16\gp^2}\, \ln \gm^2. 
}
Inserting this and the fact that the fixed point gauge couplings do
not run into \eqref{5D4DcouplingsH}, we see that we exactly reproduce 
\eqref{5DzerocouplingsH}. This means that the local fixed point
computation is consistent with the 4D zero mode result.

Of course in this example the cross check is rather easy precisely
because at the fixed points the couplings do not renormalize. In the
subsequent subsections we consider examples where the fixed point
contributions do not vanish anymore.

\subsection[Hyper multiplet on ${T^2/\Intr_3}$]{Hyper multiplet on
  $\boldsymbol{T^2/\Intr_3}$} 
\labl{sc:hyperZ3}

Now we consider a slightly less simple example of the hyper multiplet
contributions to the gauge couplings on $T^2/\Intr_3$. The basic logic
is the same as in the previous section, so we will be brief and only
emphasize the new features here. The matrix $Z_+$ in this case induces
a symmetry breaking of the form
\equ{
Z_+ = \text{diag}(\Id_{N_0}, e^{i \gf} \Id_{N_1}, e^{2i\gf} \Id_{N_2}), 
\quad 
\SU{N} \ra \SU{N_0} \times \SU{N_1} \times \SU{N_2},
} 
with $\gf = 2\gp/3$. And using the corresponding notations for the chiral
superfields that form the hyper multiplet we find the transformations 
\equ{
\gF_{+I} \ra e^{i \gf I} \, \gF_{+I}, 
\qquad  
\gF_{-I} \ra e^{i \gf (2-I)}\, \gF_{-I},
}
with $I = 0, 1,2$. From this we infer that only $\gF_{+0}$ and
$\gF_{-2}$ have zero modes. As these chiral superfields live in 
the $\rep{N}_0$ and $\crep{N}_2$ representation, respectively, the
zero mode gauge couplings renormalize as 
\equ{
\frac 1{g_0^2(\gm)} = \frac 1{g_0^2} - \frac 1{16\gp^2} \ln \gm^2, 
\qquad
\frac 1{g_1^2(\gm)} = \frac 1{g_1^2},
\qquad
\frac 1{g_2^2(\gm)} = \frac 1{g_2^2} - \frac 1{16\gp^2} \ln \gm^2.
\labl{6DzerocouplingsH}
}
In the bulk the contribution to the 6D gauge coupling is
the same as on the torus. In terms of the 4D 
renormalization scale $\gm$ we have 
\equ{
\frac {\text{Vol}_W}{g_6^2}(\gm) = \frac {\text{Vol}_W}{g_6^2} 
- \frac 2{16\pi^2} \ln \gm^2. 
}
On the fixed points, however, the
results are now more complicated than in the $\Intr_2$ case, as they
are given by \eqref{6DZNfpRenormH}. The matrix $A_+$ can be read off
from $Z_+$, hence we infer that the matrix combination in 
\eqref{6DZNfpRenormH} is given by 
\equ{
A_+ = \text{diag}(0, 1, 2) \gf, 
\qquad 
\cos A_+ + \cos (A_+ + \gf) = 
\text{diag}\Big(\frac 12, -1, \frac 12\Big).
} 
This results in the following expressions for the renormalization of
the fixed point gauge couplings 
\equ{
\frac 1{g_{f.p.0}^2(\gm)} = \frac 1{g_{f.p.0}^2} 
- \frac 1{16\gp^2} \ln \gm^2, 
\quad 
\frac 1{g_{f.p.1}^2(\gm)} = \frac 1{g_{f.p.1}^2} 
+ \frac 2{16\gp^2} \ln \gm^2, 
\quad 
\frac 1{g_{f.p.2}^2(\gm)} = \frac 1{g_{f.p.2}^2} 
- \frac 1{16\gp^2} \ln \gm^2.
}
Notice that the beta coefficient of the fixed point gauge coupling
$g_{f.p.1}$ has the opposite sign as compared to the other two fixed
point couplings.

The relation between the 4D zero mode gauge couplings,
the 6D bulk gauge coupling and the fixed point gauge couplings on
a $\Intr_3$ orbifold read 
\equ{
\frac 1{g_I^2(\gm)} = \frac 13 \frac {\text{Vol}_W}{g_6^2}(\gm) 
+ \frac 13 \frac 1{g^2_{f.p.I}(\gm)}.
\labl{6D4DcouplingsH}
}
Here we have summed over the fixed points of the $\Intr_3$ orbifold. 
Inserting the above results we find that the 4D zero mode gauge
expressions given in \eqref{6DzerocouplingsH} are indeed reproduced. In
particular, the 4D zero mode gauge coupling $g_1(\gm)$ does not
renormalize at all. Hence we see that also in this case our bulk and
fixed point results are consistent with an analysis using the zero
modes on the covering space ($T^2$) and the orbifold ($T^2/\Intr_3$).

\subsection[Vector multiplet self interactions on $\boldsymbol{S^1/\Intr_2}$]{Vector multiplet self interactions on $\boldsymbol{S^1/\Intr_2}$}
\labl{sc:NAZ2}

Both illustrations discussed above involved a single hyper multiplet
in the fundamental representation. The final two examples consider the
effects of the self interactions of the non-Abelian vector multiplet
on $\Intr_2$ and $\Intr_3$ orbifolds. We follow the same methodology
as for the hyper multiplet examples: First compute the zero mode
running and then see if it can be reproduced by combining the bulk and
fixed point couplings in the appropriate way. 

We again consider the $\Intr_2$ case on the circle for simplicity, and
take the same matrix $Z$ defined in \eqref{ZonZ2}. Writing the vector
multiplet superfields in corresponding block matrices 
\equ{
V = \pmtrx{V_0 & V_{01} \\[1ex] V_{10} & V_{1}}, 
\qquad 
S = \pmtrx{S_0 & S_{01} \\[1ex] S_{10} & S_1}, 
}  
we infer that only $V_0, V_1$ and $S_{01}, S_{10}$ have zero modes,
because of the orbifold action \eqref{NA5DORB}. Using some trace
identities to express all traces of the gauge group generators in the
fundamental representation we obtain the following zero mode beta
functions
\equ{
\frac 1{g_0^2(\gm)} = \frac 1{g_0^2} 
+ \frac{3\cdot 2N_0}{16 \pi^2} \ln \gm^2 
- \frac{2 N_1}{16 \pi^2} \ln \gm^2, 
\qquad 
\frac 1{g_1^2(\gm)} = \frac 1{g_1^2} 
+ \frac{3\cdot 2N_1}{16 \pi^2} \ln \gm^2 
- \frac{2 N_0}{16 \pi^2} \ln \gm^2. 
}
The factor of $3$ and $-1$ result, because $V_0$ and $V_1$ are 4D vector
multiplets while $S_{01}$ and $S_{10}$ are chiral multiplets.

The relation between the bulk, fixed point and zero mode gauge
couplings are as stated in \eqref{5D4DcouplingsH}. For the 5D bulk
gauge coupling we find 
\equ{
\frac{2\gp R}{g_5^2}(\gm) = \frac {2\pi R}{g_5^2} + 
\frac {2\cdot 2(N_0+N_1)}{16\gp^2} \ln \gm^2, 
}    
because the zero modes on $S^1$ are the full vector and chiral
multiplets $V$ and $S$. To compute the fixed point contributions using 
\eqref{5DZ2fpRenormNA}, we notice that by standard representation
theory  
\equ{
\tr_{\rep{Ad}} (V_0+V_1)^2 = 
2N_0 \tr_{\rep{N}_0} V_0^2 + 2N_1 \tr_{\rep{N}_1} V_1^2 + 
2N_1 \tr_{\rep{N}_0} V_0^2 + 2N_0 \tr_{\rep{N}_1} V_1^2. 
} 
In analogy to the definition of the trace $\tr_{\rep{Ad}}$ we 
define
\(
\tr_{Q,\rep{Ad}}(X Y) = 
- f_{ijk} f_{\ell mn} \get^{mj} Q^{nk} X^i Y^\ell
 \)
Using this definition and the fact that $Q$ is equal to $-1$ when it
corresponds to off-diagonal entries, we infer that
\equ{
\tr_{Q,\rep{Ad}} (V_0+V_1)^2 = 
2N_0 \tr_{\rep{N}_0} V_0^2 + 2N_1 \tr_{\rep{N}_1} V_1^2 
-2N_1 \tr_{\rep{N}_0} V_0^2 -2N_0 \tr_{\rep{N}_1} V_1^2. 
} 
Hence we find for the 4D fixed point gauge couplings 
\equ{
\frac 1{g_{f.p.0}^2(\gm)} =  \frac 1{g_{f.p.0}^2} 
+ \frac{4\cdot 2(N_0 - N_1)}{16\pi^2}  \ln \gm^2, 
\qquad 
\frac 1{g_{f.p.1}^2(\gm)} =  \frac 1{g_{f.p.1}^2} 
+ \frac{4\cdot 2(N_1 - N_0)}{16\pi^2} \ln \gm^2. 
} 
Combining these fixed point results with the bulk gauge coupling, we
see that we precisely reproduce the 4D zero mode gauge couplings.

\subsection[Vector multiplet self interactions on $\boldsymbol{T^2/\Intr_3}$]{Vector multiplet self interactions on $\boldsymbol{T^2/\Intr_3}$} 
\labl{sc:NAZ3}

Our final example discusses the non-Abelian vector multiplet self
interactions on the orbifold $T^2/\Intr_3$. Using a similar analysis
as presented above, we find that the zero modes are the vector
multiplets $V_0, V_1, V_2$ and chiral multiplets 
$S_{20}, S_{12}, S_{01}$. Consequently, the zero mode gauge couplings
read 
\equ{
\frac 1{g_{0}^2(\gm)} = \frac 1{g_0^2} + \frac 1{16\gp^2} 
\Big\{ 3\cdot 2N_0 - N_1 - N_2  \Big\} \ln \gm^2, 
} 
and cyclic permutations of the labels $0, 1, 2$. The bulk contribution
to the 4D zero mode couplings is of course the same, for the fixed
point contributions we find from \eqref{6DZNfpRenormNA} that we get 
\equ{
\frac 1{g_{f.p.0}^2(\gm)} = \frac 1{g_{f.p.0}^2} + 
\frac 1{16\pi^2} \Big\{ -14 N_0 + 7 (N_1+N_2) \Big\} \ln \gm^2,
} 
and cyclic permutations. When these results are combined we see again
that the 4D zero mode gauge couplings can be obtained from the 6D bulk
and the fixed point gauge couplings according to \eqref{6D4DcouplingsH}.

%%% Local Variables: 
%%% mode: latex
%%% TeX-master: "paper"
%%% End: 

  \section{Conclusions}
\labl{sc:concl}

In this paper we considered the renormalization of gauge kinetic
operators on orbifolds. With possible applications in string
phenomenology in mind, we focused on supersymmetric 
theories in 5D and 6D, as our results can be straightforwardly
extended to 10D string models. The $\Intr_N$ orbifolds under
investigation preserved 4D Lorentz invariance and $N=1$ supersymmetry,
which motivated us to use the language of 4D $N=1$ superfields to
describe these theories.

We presented in detail one loop computations on orbifolds using
orbifold compatible delta functions. Using this technique we computed
the gauge coupling renormalization for non-Abelian
supersymmetric gauge theories. This extended our previous work 
\cite{GrootNibbelink:2005vi} which was only concerned with Abelian
theories. For the hyper multiplet in the non-Abelian case we have
established that the renormalization of the gauge couplings at the
$\Intr_2$ fixed points is absent, but for the other $\Intr_N$
fixed points this is not the case. This result is similar to what we
had obtained in the Abelian case before. In the non-Abelian theory
there are also vector multiplet self interactions. We computed the
self energy due to these interactions on the orbifold and found that
they always give rise to renormalization both in the bulk and at all
fixed points, including the $\Intr_2$ fixed points.

In this work we performed a direct computation of the required counter
terms and therefore the renormalization at the fixed points. However,
some of our results can also be obtained indirectly by carefully
considering what happens at the zero mode level when the theory is
compactified on the orbifold or its covering space. This technique has
been advocated for example in \cite{Hebecker:2002vm}. For us this
provided an important cross check of our results for both hyper and
vector multiplet on both $\Intr_2$ and $\Intr_3$ orbifolds.

Aside from the gauge coupling renormalization, we have also
encountered some other findings. First of all in the non-Abelian case
both in 5D and 6D the local fixed point counter terms appear not to
be gauge invariant under the full bulk gauge group. It is possible to
give a non-linear completion of these terms which is invariant under
the zero mode part of the supergauge transformation. This does not
necessarily mean that gauge invariance is broken by quantum effects:
Our one loop computation only focused on two point functions. This
means that the full gauge invariant renormalized theory can only be
guessed by using arguments of supergauge invariance. In the 6D bulk
case the more or less obvious kinetic terms for the vector multiplet
have to be  completed by a Wess-Zumino-Witten term, see section
\ref{sc:class6D} and also \cite{Marcus:1983wb}. Therefore it might
not be so surprising that this could also be the case for the fixed point
contributions. What is more surprising is that this also seems to be
the case in the 5D setting. To determine these generalizations of
Wess-Zumino-Witten terms is very interesting but lies somewhat outside
the scope of the present paper. However, this question might be
interesting for future research.

The other important quantum effect is that, as in the Abelian case,
also higher derivative operators of the vector multiplet are required
in order to cancel all divergences in 6D. However, in the non-Abelian
case in 6D we also concluded that the Wess-Zumino-Witten term must
also renormalize. And in addition there must exist a higher derivative
analogue of this term, that ensures that the kinetic higher derivative
operator is gauge invariant. To determine the precise form of this
is again beyond the scope of the paper, but an interesting question
for future investigations.

\section*{Acknowledgments}

We would like to thank 
Gero von Gersdorff,
Dumitru Ghilencea, 
Arthur Hebecker, 
Hyun Min Lee
and Hans-Peter Nilles
for very useful discussions.  
The work of MH was partially supported by the EU 6th Framework
Program MRTN-CT-2004-503369 ``Quest for Unification'' and
MRTN-CT-2004-005104 ``Forces Universe''.

%%% Local Variables: 
%%% mode: latex
%%% TeX-master: "paper"
%%% End: 

%\renewcommand{\theequation}{\Alph{thechapter}.\arabic{equation}}
%\renewcommand{\theequation}{\Alph{\thesection}.\arabic{equation}}
\appendix 
\def\theequation{\thesection.\arabic{equation}}
\setcounter{equation}{0}
%\labl{sc:RegInt}

\section{Results for the Feynman graphs}
\labl{sc:graphs_expr}

Here we give the results for the Feynman graphs that have been
calculated. All results are formulated in 6D notation
on the orbifold $T^2/\Intr_N$. These results can also be applied to the
orbifold $S^1/\Intr_2$:  The 5D situation is obtained when one
replaces $z=y$ and neglects all dependence on $\bz$, the derivatives
change as $\del = \bdel = \del_5$, and instead of the six dimensional
orbifold compatible delta functions (\ref{6D_delta_functions}) one
uses their five dimensional counterparts
(\ref{5D_delta_functions}).

\subsection{Gauge multiplet corrections to the vector multiplet self energy}
\labl{sc:AppGauge}

In this appendix we give the superspace and the resulting scalar
integral for the gauge multiplet contributions to the vector multiplet
corresponding to the diagrams in figure \ref{fg:VV_from_VSC}. 
We use the labels for these diagrams suggested by that figure to refer
to the contributions of the various topologies and superfields. 
%%% V-[S]-V loop %%%
 \equl{
\text{\ref{fg:VV_from_VSC}.A}      =  f_{ijk}f_{\ell mn} \, \int (d^6x)_{12}\;d^4\theta \Big[
      V_1^i \Box P_0 V_2^\ell \,\,\, \propsix{1}{2} \;
      \dtwup{6(\bS)}{21}{m}{j} \propsix{1}{2} \;
      \dtwup{6(S)}{21}{n}{k} + 
\\[2ex] 
   - 2 \, V_1^i V_2^\ell \,\,\, \propsix{1}{2} \; \dtwup{6(\bS)}{21}{m}{j}
      \propsix{\Box_2}{2} \; \dtwup{6(S)}{21}{n}{k} \Big]. 
\labl{Afinal}
\qquad 
}
%%% V-[V]-V loop %%%
\equl{
\text{\ref{fg:VV_from_VSC}.B} 
=  f_{ijk}f_{\ell mn} \, \int (d^6x)_{12}\;d^4\theta\,V_1^i
\Box \left(\frac 12 P_+ + \frac 12 P_- - \frac 52 P_0 \right)
V_2^\ell
\; \times \\[2ex] \times \; 
\propsix{1}{2}\, \dtwup{6(V)}{21}{m}{j} \propsix{1}{2}
\; \dtwup{6(V)}{21}{n}{k}.  
\qquad 
}
%%% V-[V/S]-V loop %%%
\equl{
\text{\ref{fg:VV_from_VSC}.C}      =  f_{ijk}f_{\ell mn} \, \int (d^6x)_{12}\;d^4\theta \Big[ \del
      V_1^i \bdel V_2^\ell \,\,\, \propsix{1}{2} \;
      \dtwup{6(V)}{21}{m}{j} \propsix{1}{2} \;
      \dtwup{6(\bS)}{21}{n}{k} + 
\\[2ex]  
      - \del V_1^i V_2^\ell \,\,\, \propsix{\bdel_2}{2} \;
      \dtwup{6(V)}{21}{m}{j} \propsix{1}{2} \;
      \dtwup{6(\bS)}{21}{n}{k}
      -  V_1^i \bdel V_2^\ell \,\,\, \propsix{\del_1}{2} \;
      \dtwup{6(V)}{21}{m}{j} \propsix{1}{2} \;
      \dtwup{6(\bS)}{21}{n}{k} + 
\\[2ex]  
       +   V_1^i V_2^\ell \,\,\, \propsix{\del_1\bdel_2}{2} \;
       \dtwup{6(V)}{21}{m}{j} \propsix{1}{2} \;
       \dtwup{6(\bS)}{21}{n}{k} \Big]. 
\qquad 
}
%%% V-[C]-V loop %%%
\equl{
\text{\ref{fg:VV_from_VSC}.D}      
=  f_{ijk}f_{\ell mn} \, \int (d^6x)_{12}\;d^4\theta \Big[ 
2\, V_1^i V_2^\ell \,\,\, \propsix{1}{2} \; \dtwup{6(V)}{21}{m}{j}
\propsix{\Box_2}{2} \; \dtwup{6(V)}{21}{n}{k} + 
\\[2ex] 
-\frac{1}{2} \, V_1^i \Box \left( P_+ + P_- + P_0
 \right) V_2^\ell \,\,\, \propsix{1}{2} \; \dtwup{6(V)}{21}{m}{j}
 \propsix{1}{2} \; \dtwup{6(V)}{21}{n}{k} \Big]. 
\qquad
}
%%% V-[V]-V tadpole %%%
\equl{
\shoveright{
\text{\ref{fg:VV_from_VSC}.E} = -\frac{1}{3} f_{ijk}f_{\ell mn} \eta^{nk} \int
  (d^6x)_{12}\;d^4\theta\,V_1^i  V_2^\ell\,
%\propsix{(\Box+\del\bdel)_2}{2}
\, \dtw{6(V)}{21}{a}{j}
  \propsix{\eta^{ab}}{2} \; \dtw{6(V)}{21}{b}{m}. 
} 
}
%%% V-[S]-V tadpole %%%
\equl{
\shoveright{
\text{\ref{fg:VV_from_VSC}.F}  =  
2 \, f_{ijk}f_{\ell mn} \, \eta^{nk} \int (d^6x)_{12} \;
       d^4\theta \, V_1^i V_2^\ell \,\,
%       \propsix{(\Box+\del\bdel)_2}{2} \; 
\dtw{6(\bS)}{21}{a}{m} \;
       \propsix{\eta^{ab}}{2} \; \dtw{6(S)}{21}{b}{j}. 
}
}
%%% V-[C]-V tadpole %%%
\equl{ 
\shoveright{
\text{\ref{fg:VV_from_VSC}.G}  =   -\frac{2}{3} \, f_{ijk}f_{\ell mn} \, \eta^{nk} \int
    (d^6x)_{12} \; d^4\theta \, V_1^i V_2^\ell \,\,
%    \propsix{(\Box+\del\bdel)_2}{2} \, 
\dtw{6(V)}{21}{a}{j} \;
    \propsix{\eta^{ab}}{2} \, \dtw{6(V)}{21}{b}{m} . 
}
}
The diagrams in figure \ref{fg:SS_from_VS} give rise to the  $S \bS$
and $V S$ self energies. Written in terms of two orbifold
compatible delta functions we have:
%%% S-[V/S]-\bS loop %%%
\equl{
\shoveright{
\text{\ref{fg:SS_from_VS}.A}  = ~ 2 \, f_{ijk}f_{\ell mn} \, \int (d^6x)_{12}\;d^4\theta \, S_1^i \bS_2^\ell \,\,
    \propsix{1}{2} \; \dtwup{6(V)}{21}{m}{j}
    \propsix{1}{2} \; \dtwup{6(\bS)}{21}{n}{k}. 
 }
}
%%% V-[V/S]-\bS loop %%%
\equl{
\text{\ref{fg:SS_from_VS}.B}  =  -\sqrt{2} \, f_{ijk}f_{\ell mn} \, \int (d^6x)_{12}\;d^4\theta \, \Big( 2\, \del V_1^i \bS_2^\ell \,\,
    \propsix{1}{2} \; \dtwup{6(V)}{21}{m}{j}
    \propsix{1}{2} \; \dtwup{6(\bS)}{21}{n}{k} +
\\[2ex]
 + V_1^i \bS_2^\ell \,\,
    \propsix{1}{2} \; \dtwup{6(V)}{21}{m}{j}
    \propsix{\del_1}{2} \; \dtwup{6(\bS)}{21}{n}{k} \Big).
\qquad  
}
%%% V-[C]-\bS loop %%%
\equl{
\shoveright{
\text{\ref{fg:SS_from_VS}.C}  =  +\sqrt{2} \, f_{ijk}f_{\ell mn} \,
\int (d^6x)_{12}\;d^4\theta \, V_1^i \bS_2^\ell \,\, 
    \propsix{1}{2} \; \dtwup{6(V)}{21}{m}{j}
    \propsix{\del_1}{2} \; \dtwup{6(\bS)}{21}{n}{k}.
 }
}

\subsection{Hyper multiplet corrections to the vector multiplet self energy}
\labl{sc:AppHyper}

In this appendix we give the scalar integral 
expression for the supergraphs given in figures
\ref{fg:VV_from_hypers} and \ref{fg:SS_from_Hypers} at the level of two remaining orbifold
compatible delta functions. 
%%% T_{++} loop %%%
\equl{
\text{\ref{fg:VV_from_hypers}.A}_\pm = 
 2 \int \left( d^6x \right)_{12} d^4\gth \; \tr \Big[ V_1
    \propsix{1}{2} \tgd_{21}^{6(+)} \Big( 
 \frac{1}{2} \Box P_0  V_2 + V_2 \Box_2 \Big)
    \propsix{1}{2} \tbgd_{21}^{6(+)} +
\\[2ex]
 + V_1
    \propsix{1}{2} \tbgd_{21}^{6(-)} \Big( 
 \frac{1}{2} \Box P_0  V_2 + V_2 \Box_2 \Big)
    \propsix{1}{2} \tgd_{21}^{6(-)} \Big]. 
\qquad 
}
%%% T_{+-} loop %%%
\equl{
\shoveright{
\text{\ref{fg:VV_from_hypers}.B} = ~ 4 \int \left( d^6x \right)_{12} d^4\gth \; \tr \left[
V_1 \propsix{\del_1}{2} \tbgd_{21}^{6(+)} V_2
    \propsix{\bdel_2}{2} \tgd_{21}^{6(-)} \right]. 
}
}
%%% T_+ tadpole %%%
\equl{
\shoveright{
\text{\ref{fg:VV_from_hypers}.C}_\pm = 
 -2 \int \left( d^6x \right)_{12} d^4\gth \; \tr \Big[ V_1 V_2
%  \propsix{(\Box+\del\bdel)_2}{2}
 \tgd_{21}^{6(+)} \frac{1}{(\Box+\del\bdel)_2} \tbgd_{21}^{6(+)} 
%+\non \\[2ex]& 
 -2 V_1 V_2
%  \propsix{(\Box+\del\bdel)_2}{2} \tgd_{21}^{6(-)} 
\frac{1}{(\Box+\del\bdel)_2} \tbgd_{21}^{6(-)} \Big].  
}
}
For the $S\bS$ and $S V$ self energies we find from figure
\ref{fg:SS_from_Hypers}: 
%%% S-[+/-]-\bS loop %%%
\equl{
\shoveright{
\text{\ref{fg:SS_from_Hypers}.A}_\pm = ~ 2 \int \left( d^6x
\right)_{12} d^4\gth \; \tr \Big[ S_1 
  \propsix{1}{2} \tgd_{21}^{6(+)} \bS_2 \propsix{1}{2}
  \tgd_{21}^{6(-)} \Big].
}
}
%%% V-[+/-]-\bS loop %%%
\equl{
\shoveright{
\text{\ref{fg:SS_from_Hypers}.B}_\pm = -2\sqrt{2} \int \left( d^6x
\right)_{12} d^4\gth \; \tr \Big[ \del V_1 
  \propsix{1}{2} \tgd_{21}^{6(+)} \bS_2 \propsix{1}{2}
  \tgd_{21}^{6(-)} \Big]. 
}
}

%%% Local Variables: 
%%% mode: latex
%%% TeX-master: "paper"
%%% End: 

  \setcounter{equation}{0}
  \section{Fourier transformation conventions}
\labl{sc:fourier}

In the main text we need to perform Fourier transformations between
coordinate and momentum space. We describe our conventions for the 6D
situation compactified on the torus $T^2$. Because of the notation in
6D and 5D introduced in sections \ref{sc:S1Z2} and \ref{sc:T2ZN},
respectively, the reduction to the 5D integrals on the circle $S^1$ is
straightforward. We define the Fourier transformations as  
\be
  \label{fourier}
  A (x,z) = \int \frac{d^dp}{(2\pi\, \gm)^d} \sum_{n \in \gL_K} A(p,n)
  e^{i\left( px + nz + \bn\bz \right)},  
\ee
and
\be
  A (p,n) = \frac{2 \gm^d}{\text{Vol}_W} \int d^dx d^2z A(x,z)
  e^{-i\left( px + nz + \bn\bz \right)}. 
\ee
We have introduced the regularization scale $\gm$ such that the 
coordinate and momentum Fourier transforms have the same mass
dimension. The coordinate space delta function is given by
\be
  \gd^d (x_2-x_1) \gd^2 (z_2-z_1) = \int \frac{d^dp}{(2\pi)^d}
  \frac{1}{\Vol_W} \sum_{n \in \gL_K} e^{i\left( p(x_2-x_1) +
      n(z_2-z_1) + \bn(\bz_2-\bz_1) \right)}. 
\ee
The delta function in momentum space can be expanded as
\be
  \gd^d (p_2-p_1) \gd^2 (n_2-n_1) = 2 \int \frac{d^dx d^2z}{(2\pi)^d
    \text{Vol}_W} e^{i \left (p_2-p_1)x + (n_2-n_1)z +
      (\bn_2-\bn_1)\bz \right)}. 
\ee

  \setcounter{equation}{0}
  \section{Theta functions}
  \labl{sc:theta_functions}

 \subsection{Genus one theta functions}

The genus one theta function on the Kaluza Klein lattice is defined as
\be
  \label{g1th_K}
  \gth_K \brkt{\ga}{\gb} (\gs|\gt) = \sum_{n \in \Intr/R}
  e^{i\frac{\gt}{2} (n-\ga)^2 + i\left[ (\gs-\gb)(n-\ga)\right]}.
\ee
The theta function is translation invariant under a shift of $\ga$
by an element of the Kaluza Klein lattice or a shift of $\gb$ by an
element of the winding mode lattice
\be
  \label{th_fct_prop1}
  \gth_K \brkt{\ga+n}{\gb} (\gs|\gt) = \gth_K \brkt{\ga}{\gb}
  (\gs|\gt), \quad \quad n\,\in\,\gL_K, 
\ee
\be
  \label{th_fct_prop2}
  \gth_K \brkt{\ga}{\gb+w} (\gs|\gt) = \gth_K \brkt{\ga}{\gb}
  (\gs-w|\gt), \quad \quad w\,\in\,\gL_W. 
\ee
The genus one theta function on the winding mode lattice is defined as
\be
  \label{g1th_W}
  \gth_W \brkt{\ga}{\gb} (\gs|\gt) = \sum_{w \in 2\pi R} e^{i
    \frac{\gt}{2} (w-\ga)^2 + i (w-\ga)(-\gb)}. 
\ee
The relation between $\gth_K$ and $\gth_W$ is
\be
  \gth_K \brkt{\ga}{\gb} (\gs|\gt) = R \, \sqrt{\frac{2\pi}{-i\gt}} \,
  e^{-\frac{i}{2\gt}\gt^2 + i\ga\gb} \; \gth_W\brkt{\gb}{-\ga}
  \left(\frac{-\gs}{\gt}\left|\frac{-1}{\gt}\right.\right).
\ee
This can be obtained by using Poisson resummation, which allows us to
rewrite a complex exponential function that is summed over the
Kaluza-Klein lattice $\gL_K$ into a delta function that is summed over
the winding mode lattice and vice versa. Concretely, we have  
\be
  \frac{1}{2\pi R} \sum_{n \in \gL_K} e^{iny} = \sum_{w \in \gL_W}
  \gd(y-w), \quad \quad y\,\in\,\Real  
\ee
and
\be
  R \sum_{w \in \gL_W} e^{iwp} = \sum_{n \in \gL_K} \gd(p-n), \quad
  \quad p\,\in\,\Real. 
\ee

  \subsection{Genus two theta functions}

The genus two theta function on the Kaluza-Klein lattice $\gL_K$ is
defined by 
\be
  \label{g2th_K}
  \gth_K \brkt{\ga}{\gb} (\gs|\gt) = \sum_{n \in \gL_K} e^{i
    \frac{\gt}{2} |n-\ga|^2 + i \left[ (\bgs-\bgb)(\bn-\bga) +
      (\gs-\gb)(n-\ga) \right]}.
\ee
Also the genus two theta function fulfills the translation invariance
properties (\ref{th_fct_prop1}) and (\ref{th_fct_prop2}). The genus
two theta function on the winding mode lattice is defined by 
\be
  \label{g2th_W}
  \gth_W \brkt{\gb}{\ga} (\gs|\gt) = \sum_{w \in \gL_W} e^{2i\gt
    |w-\gb|^2 + i(\bw-\bgb)(2\gs-\bga) + (w-\gb)(2\bgs-\ga)}. 
\ee
The relation between $\gth_K$ and $\gth_W$ is given by 
\be
  \gth_K \brkt{\ga}{\gb}\left(\gs|\gt\right) = \Vol \;
  \frac{2\pi}{-i\gt} \; e^{i \left( -\frac{2}{\gt}|\gs|^2 + \ga\gb
      +\bga\bgb \right)} \; \gth_W
  \brkt{\gb}{-\ga}\left(\frac{-\gs}{\gt}\left|\frac{-1}{\gt}\right.\right). 
\ee
This is obtained by Poisson resummation on the torus 
\be
  \frac{1}{\text{Vol}_W} \sum_{n \in \gL_K} e^{i(nz+\bn\bz)}
 = \sum_{w \in \gL_W} \frac{1}{2} \, \gd^2 (z-w), \quad \quad
 z\,\in\,\Cplx,   
\ee
\be
  \frac{1}{\text{Vol}_K} \sum_{w \in \gL_w} e^{i(wp+\bw\bp)} 
= \sum_{n  \in \gL_K} 2 \, \gd^2 (p-n), \quad \quad p\,\in\,\Cplx, 
\ee
where
\be
  \gd^2 (p) = \frac{1}{2} \, \gd (p_5) \gd (p_6), \quad \quad \gd^2(z) 
= 2 \, \gd (x_5) \gd (x_6). 
\ee
Therefore, in the case $\gt = \frac{2it}{\gm^2}$, $\ga = sl$, and 
$\gs = \gb = 0$
\be
  \label{poisson_resum}
  \gth_K \brkt{sl}{0} \left(0|\mbox{$\frac{2it}{\gm^2}$}\right) =  \; \Vol \left(
    \frac{\pi\gm^2}{t} \right)^{\frac{D-d}{2}} \gth_W \brkt{0}{-sl}
  \left(0|\mbox{$\frac{i\gm^2}{2t}$}\right),  
\ee
holds both for the theta functions on the circle and on the torus.

%%% Local Variables: 
%%% mode: latex
%%% TeX-master: "paper"
%%% End: 

  \setcounter{equation}{0}
  \section{Regularization of the momentum integral}
  \labl{Regularization_Integral}

In order to determine the counterterms, we need to calculate the
divergent part of the integral 
\be
  \label{divergent_integral}
  I_{D} = i \int \frac{d^dp_E}{(2\pi)^d} \frac{1}{\Vol_W} \sum_n
  \frac{1}{p^2 + |n|^2 + m^2}\frac{1}{(p-k)^2 + |n-l|^2 + m^2}, 
\ee
which is obtained after a Wick rotation. We can replace the
integration over the volume of the continuous momenta by the
integration over the radius 
\be
  \int \frac{d^dp_E}{(2\pi)^d} = \frac{2 (\gm)^{4-d}}{(4\pi)^{d/2}
    \, \gG \left( {\frac{d}{2}} \right)} \int_0^\infty dp \; p^{d-1}.
\ee
The non-compact 4D integral is extended to $d= 4 -2 \ge$ dimensions
using the standard procedure of dimensional regularization of scalar
integrals. Furthermore we use the identity
\be
  \frac{1}{M_i^2} = \frac{1}{\gm^2} \int_0^\infty dt \; e^{-t M_i^2/\gm^2}
\ee
where $M_i$ are the momentum-dependent denominators of
(\ref{divergent_integral}). With the help of a Feynman parameter $s$
the integral (\ref{divergent_integral}) can be written as 
\be
  I_{D} = i \frac{1}{(4\pi)^{d/2}}
  \frac{1}{(2\pi)^{D-d}\;\text{Vol}} \int_0^1 ds \int_0^\infty
  \frac{dt}{t^{d/2-1}} e^{-t \left[ s(1-s)(k^2+|l|^2) + m^2
    \right]/\gm^2 } \gth_K \brkt{sl}{0} \left(0|\mbox{$\frac{2it}{\gm^2}$}\right)
  \ee
where for $D-d = 1$, $2$ we take for $\gth_K$ the $\gth$ function of
the circle or the torus, defined in (\ref{g1th_K}) and (\ref{g2th_K}),
respectively. After application of the Poisson resummation formula
(\ref{poisson_resum}) which is valid both on the circle and on the
torus, we obtain 
\be
  I_{D} = i \frac{\gm^{D-d}}{(4\pi)^{D/2}} \int_0^1 ds \int_0^\infty
  \frac{dt}{t^{D/2-1}} e^{-t \left[ s(1-s)(k^2+|l|^2) + m^2
    \right]/\gm^2 } \gth_W \brkt{0}{-sl}
  \left(0|\mbox{$\frac{i\gm^2}{2t}$}\right) 
\ee
Because $\gth_W-1$ cannot lead to UV divergencies, 
we can put $\gth_W = 1$ in order to determine the divergent part.  
We find 
\be
  I_{D}\big|_{\text{div}} = i \frac{1}{\gm^d} \left(
    \frac{\gm^2}{m^2} \right)^2 \left( \frac{m^2}{4\pi}
  \right)^{\frac{D}{2}} \sum_{n \geq 0} (-)^n \frac{\gG \left(
      n+2-\frac{D}{2} \right) \gG(n+1)}{\gG(2n+2)} \left(
    \frac{K^2}{m^2} \right)^n. 
\ee
In the 6D case we obtain 
\be
  \label{divergence_6D}
  I_{6-2\ge}\big|_{\text{div}} =  \frac{i}{(4\pi)^3} \Big[
    \Big(\frac{1}{\bge} + \ln \frac{\gm^2}{m^2} \Big) 
\Big( m^2 + \frac{1}{6} (k^2 + |l|^2) \Big) +
    m^2 \Big], 
\ee
where we have defined $\frac{1}{\bge} = \frac{1}{\ge} - \gg + \ln
4\gp$. Here only the terms with $n\;\in\;\{0,1\}$ contribute to the
divergent part and we have neglected terms with higher $n$. In the
5D case the expression reads 
\be
  \label{divergence_5D}
  I_{5-2\ge}\big|_{\text{div}} = -i \frac{1}{(4\pi)^2} |m|,
\ee
where only the $n=0$ term has been taken into account. The four
dimensional case can also be traced back when one neglects the
summation $\frac{1}{\Vol_W} \sum_n$. This results in 
\be
  \label{divergence_4D}
  I_{4-2\ge}\big|_{\text{div}} = \frac{i}{(4\pi)^2} \Big(
    \frac{1}{\ge} -\gg + \ln 4 \gp + \ln \frac{\gm^2}{m^2} 
  \Big). 
\ee

%%% Local Variables: 
%%% mode: latex
%%% TeX-master: "paper"
%%% End: 

\bibliographystyle{paper}
{\small
\bibliography{paper}
}

\end{fmffile}
\end{document}